\documentclass[]{article}
\usepackage{authblk}

\usepackage[utf8]{inputenc} 
\usepackage[T1]{fontenc}    
\usepackage{hyperref}       
\usepackage{url}            
\usepackage{booktabs}       
\usepackage{amsfonts}       
\usepackage{nicefrac}       
\usepackage{microtype}      
\usepackage{fullpage}

\usepackage{adjustbox}
\usepackage{hyperref}
\usepackage{url}
\usepackage{graphicx}
\usepackage{amsmath}
\usepackage{bbm}
\usepackage{paralist}
\usepackage{Definitions}
\usepackage[ruled,noline]{algorithm2e} 
\usepackage{color}
\usepackage{enumitem}
\usepackage{mathtools}
\usepackage{subfig}
\usepackage{graphicx}
\usepackage{tabularx}
\usepackage[dvipsnames]{xcolor}
\usepackage{stackengine}
\usepackage{booktabs}
\usepackage{hhline}
\usepackage[multiple]{footmisc}

\newcommand{\xhdr}[1]{\vspace{0.1mm}\noindent{{\bf #1. }}}
\newcommand{\xhdrq}[1]{\vspace{0.1mm}\noindent{{\bf #1}}}

\newcommand{\explain}[2]{\underset{\mathclap{\overset{\uparrow}{#2}}}{#1}}
\newcommand{\explainup}[2]{\overset{\mathclap{\underset{\downarrow}{#2}}}{#1}}

 \graphicspath{{./FIG/}}

\newcommand{\denselist}{ \itemsep -3pt\topsep-20pt\partopsep-20pt }

\title{Distilling Information Reliability and Source Trustworthiness from Digital Traces}

\author[1]{Behzdad Tabibian}
\author[2]{Isabel Valera}
\author[3]{Mehrdad Farajtabar}
\author[3]{Le Song}
\author[1]{\\Bernhard Sch\"{o}lkopf}
\author[2]{Manuel Gomez-Rodriguez}

\affil[1]{Max Planck Institute for Intelligent Systems, me@btabibian.com, bs@tue.mpg.de}
\affil[2]{Max Planck Institute for Software Systems, ivalera@mpi-sws.org, manuelgr@mpi-sws.org}
\affil[3]{Georgia Institute of Technology, mehrdad@gatech.edu, lsong@cc.gatech.edu}

\date{}

\begin{document}


\maketitle

\begin{abstract}
Online knowledge repositories typically rely on their users or dedicated editors to evaluate the reliability of their content.
These evaluations can be viewed~as noi\-sy measurements of both information reliability and information
source trustworthiness.
Can we leverage these noisy evaluations, often biased, to distill a robust, unbiased and interpretable measure of both notions?

In this paper, we argue that the \emph{temporal traces} left by these noisy evaluations give cues on the reliability of the information
and the trustworthiness of the sources.
Then, we propose a temporal point process modeling framework that links these temporal traces to robust,
unbiased and interpretable notions of information reliability and source trustworthiness.
Furthermore, we develop an efficient convex optimization procedure to learn the pa\-ra\-me\-ters of the model from historical traces.
Experiments on real-world data ga\-thered from \textit{Wikipedia} and \textit{Stack Overflow} show that our modeling framework accurately predicts evaluation events, provides an interpretable
measure of information reliability and source trustworthiness, and yields interesting insights about real-world events.
\end{abstract}

\section{Introduction}
\label{sec:intro}
Over the years, the Web has become a vast re\-po\-si\-to\-ry of information and knowledge about a rich variety of to\-pics and real-world events -- one much
larger than anything we could hope to accumulate in conventional textbooks and traditional news media outlets.
Unfortunately, due to its immediate nature, it also contains an ever-growing number of opinionated, inaccurate or false facts, urban
legends and unverified stories of unknown or questionable origin, which are often refuted over time.\footnote{\scriptsize \url{http://www.snopes.com}}\footnote{\scriptsize \url{http://www.factcheck.org}}
To overcome this problem, online knowledge repositories, such as \textit{Wikipedia},  \textit{Stack Overflow} and \textit{Quora}, put in place different evaluation mechanisms to increase
the reliability of their content. These mechanisms can be typically classified as:
\begin{itemize}
\denselist
\item[I.] {\bf Refutation:} A user refutes, challenges or questions a statement contributed by another user or a piece of content originated from an external web source. For example, in \textit{Wikipedia}, an editor can refute a questionable, false or incomplete statement in an article by removing it.
\item[II.] {\bf Verification:} A user verifies, accepts or supports a statement contributed by another user or a piece of content originated from an external web source. For
example, in  \textit{Stack Overflow}, a user can accept or up-vote the answers provided by other users.
\end{itemize}
However, these evaluation mechanisms only provide noisy measurements of the
reliability of information and the trustworthiness of the information sources.
Can we leverage these noisy measurements, often biased, to distill a robust, unbiased
and interpretable measure of both notions?

In this paper, we argue that the \emph{temporal traces} left by these noisy evaluations give
cues on the reliability of the information and the trustworthiness of the sources.
For example, while statements provided by an untrustworthy user will be often spotted by other users as unreliable and refuted quickly, statements provided by trustworthy users will be refuted less often.
However, at a particular point in time, a statement about a complex, controversial or time evol\-ving topic, story, or more generally, \emph{knowledge item}, may be
refuted by other users independently of the source. In this case, quick refutations will not reflect the trustworthiness of the source but the intrinsic unreliability of the knowledge
item the statement refers to.

To explore this hypothesis, we propose a temporal point process modeling framework of refutation and verification in online knowledge re\-po\-si\-to\-ries, which leverages the
above mentioned temporal traces to obtain a meaningful measure of both information reliability and source trustworthiness.
The key idea is to disentangle to what extent the temporal information in a statement evaluation (verification or refutation) is due to the intrinsic unreliability of the involved knowledge item, or to the trustworthiness of the source providing the statement.
To this aim, we model the times at which statements are added  to a knowledge item as a coun\-ting process, whose intensity captures the temporal evolution of  the reliability of the  item---as a knowledge item becomes
more reliable, it is less likely to be modified.
Moreover, each added statement is supported by an information source and evaluated by the users in the knowledge repository at some point after its addition time. Here, we model the evaluation time of each statement as
a survival process, which starts at the addition time of the statement and whose intensity captures both the trustworthiness of the associated source and the unreliability of the involved knowledge item.

For the proposed model, we develop an efficient method to find the optimal model parameters that jointly maximize the likelihood of an observed set of statement addition and
evaluation times.
This efficient algorithm allows us to apply our framework to $\sim$$19$ million addition and refutation events in $\sim$$100$ thousand \textit{Wikipedia} articles and $\sim$$1$ million addition and verification events in $\sim$$378$ thousand questions in \textit{Stack Overflow}.
Our experiments show that our model accurately predicts whether a statement in a \textit{Wikipedia} article (an answer in  \textit{Stack Overflow}) will be refuted (verified), it provides interpretable measures of source trustworthiness and
information reliability, and yields interesting insights:\footnote{\scriptsize We will release code for our inference method, datasets and a web interface for exploring results at \url{http://btabibian.com/projects/reliability}.}
\begin{itemize}
\denselist
\item[I.] Most active sources are generally more trustworthy, however, trustworthy sources can be also found among less active ones.
\item[II.] Changes on the reliability of a \textit{Wikipedia} article over time, as inferred by our framework, match external noteworthy events.
\item[III.] Questions and answers in  \textit{Stack Overflow} cluster into groups with similar levels of difficulty and popularity.
\end{itemize}

\xhdr{Related work}
The research area most closely related to ours is on truth discovery and source trustworthiness.
The former aims at resolving conflicts among noisy information published by diffe\-rent sources and the
latter assesses the quality of a source by means of its ability to provide correct factual information.
Most previous works have studied both problems together and measure the trustworthiness of a source
using link-based measures~\cite{borodin2005link,gyongyi2004combating}, information retrieval based measures~\cite{wu2007corroborating},
accuracy-based measures~\cite{dong2014data,dong2015knowledge,xiao2016towards}, content-based measures~\cite{adler2007content},
and graphical model analysis~\cite{pasternack2013latent,yin2011semi,zhao2012probabilistic,zhao2012bayesian}.
A recent line of work~\cite{li2015discovery,liu2011online,pal2012information,wang2014towards} also considers scenarios in which the truth may change over time.
However, previous work typically shares one or more of the following limitations, which we address in this work:
(i) they only support knowledge triplets (subject, predicate, object) or structured knowledge; 
(ii) they assume there is a \emph{truth}, however, a statement may be under discussion when a source writes about it; and,
(iii) they do not distinguish between the unreliability of the knowledge item to which the statement refers and the trustworthiness of the source.

Temporal point processes have been previously used to model information cascades~\cite{gomez11netrate,du13nips,daneshmand14icml}, social activity~\cite{farajtabar2016multistage,smart16,shaping14nips}, badges~\cite{rtpp16}, network evolution~\cite{hunter2011dynamic,farajtabar2015coevolve}, opinion dynamics~\cite{de2015modeling}, or
product competition~\cite{competing15icdm}. However, to the best of our knowledge, the present work is the first that leverages temporal point processes in the context
of information reliability and source trustworthiness.

\section{Background on Temporal Point Processes}
\label{sec:background}
A temporal point process is a stochastic process whose rea\-li\-za\-tion consists of a list of discrete events localized in time, $\cbr{t_i}$ with $t_i \in \RR^+$ and $i \in \ZZ^+$.
Many different types of data produced in social media and the Web can be represented as temporal point processes~\cite{de2015modeling,farajtabar2015coevolve,competing15icdm}.
A temporal point process can be equivalently re\-pre\-sen\-ted as a counting process, $N(t)$, which records the number of events up to time $t$,
and can be characterized via its conditional intensity function --- a stochastic model for the time of the next event given all the times of previous events. More formally, the
conditional intensity function $\lambda^*(t)$ (intensity, for short) is given by
\begin{align}
  \lambda^*(t)dt := \PP\cbr{\text{event in $[t, t+dt)$}|\Hcal(t)} = \EE[dN(t) | \Hcal(t)], \nonumber
\end{align}
where $dN(t) \in \cbr{0,1}$ denotes the increment of the process, $\Hcal(t)$ denotes the history of event times $\cbr{t_1, t_2, \ldots, t_n}$ up to but not in\-clu\-ding time $t$, and the
sign $^{*}$ indicates that the intensity may depend on the history.
Then, given a time $t_i \geqslant t_{i-1}$, we can also characterize the conditional probability that no event happens during $[t_{i-1}, t_i)$ and the conditional density that an
event occurs at time $t_i$ as $S^*(t_i) = \exp(-{\scriptsize \int_{t_{i-1}}^{t_i}} \lambda^*(\tau) \, d\tau)$ and $f^*(t_i) = \lambda^*(t_i)\, S^*(t_i)$, respectively.
Furthermore, we can express the log-likelihood of a list of events $\cbr{t_1,t_2,\ldots,t_n}$ in an observation window $[0, T)$ as \cite{AalBorGje08}
\begin{align}
  \label{eq:loglikehood_fun}
  \Lcal = \sum_{i=1}^n \log \lambda^*(t_i) - \int_{0}^T \lambda^*(\tau)\, d\tau.
\end{align}
This simple log-likelihood will later enable us to learn the parameters of our model from observed data.
Finally, the functional form of the intensity $\lambda^*(t)$ is often designed to capture the phenomena of interests. Some useful functional forms we will use later
are~\cite{AalBorGje08}:
\begin{itemize}
\item[I.] {\bf Poisson process.} The intensity is assumed to be independent of the history $\Hcal(t)$, but it can be a time-varying function,~\ie, $\lambda^*(t) = g(t)\geqslant 0$;
\item[II.] {\bf Hawkes Process.} The intensity models a mutual excitation between events,~\ie,
 \begin{equation}
   \label{eq:hawkes}
   \lambda^*(t) 
                = \mu + \alpha \sum\nolimits_{t_i \in \Hcal(t)} \kappa_{\omega}(t-t_i),
 \end{equation}
 where $\kappa_{\omega}(t)$ is the triggering kernel, $\mu\geqslant 0$ is a baseline intensity independent of history.
 Here, the occurrence of each historical event increases the intensity by a certain amount determined by the kernel and the weight $\alpha \geqslant 0$, making the intensity history dependent
 and a stochastic process by itself; and,
\item[III.] {\bf Survival process.} There is only one event for an instantiation of the process,~\ie,
 \begin{align}
   \label{eq:survival_process}
   \lambda^*(t) = g(t)(1 - N(t)),
 \end{align}
 where $\lambda^*(t)$ becomes $0$ if an event already ha\-ppened before $t$ and $g(t) \geqslant 0$.
\end{itemize}

\section{Proposed Model}
\label{sec:formulation}
In this section, we formulate our modeling framework of verification and refutation in knowledge repositories, star\-ting with the data representation it uses.

\xhdr{Data representation}
The digital traces generated du\-ring the construction of a knowledge repository can be re\-pre\-sen\-ted using the following three entities:
the \emph{statements}, which are associated to particular \emph{knowledge items}, and the \emph{information sources}, which support each of the
statements. More specifically:

\emph{--- An information source} is an entity that supports a statement in a knowledge repository, \ie, the web source an editor uses to
support a paragraph in \textit{Wikipedia}, the user who posts an answer on a Q\&A site, or the software de\-ve\-lo\-per who contributes a piece of code in \textit{Github}.
We denote the set of information sources in a knowledge repository as $\Scal$.

\emph{--- A statement} is a piece of information contributed to a knowledge repository, which is characterized by its addition time $t$, its evaluation time $\tau$, and the
information source $s \in \Scal$ that supports it. Here, we represent each statement as the triplet
  \begin{align} \label{eq:statement}
    e = (\explainup{s}{\text{source}}, ~~~~\explain{t}{\text{addition time}}, ~~~~\explainup{\tau}{\text{evaluation time}}),
  \end{align}
where an evaluation may correspond either to a verification or refutation.\footnote{\scriptsize For clarity, we assume that a knowledge repository either uses refutation or
verification. However, our model can be readily extended to knowledge repositories using both.}
Moreover, if a statement is \emph{never} refuted or verified, then we set $\tau = \infty$.

\emph{---  A knowledge item} is a collection of statements. For example, a knowledge item corresponds to an article in \textit{Wi\-ki\-pe\-dia}; to a question and its answer(s) in a Q\&A site; or
to a software project on \textit{Github}. Here, we gather the history of the $d$-th knowledge item, $\Hcal_d(t)$, as the set of statements added to the knowledge item $d$ up to but not including time $t$, \ie,
  \begin{equation} \label{eq:history}
  \Hcal_d(t) = \{e_i | t_i<t\}.
  \end{equation}

In most knowledge repositories, one can recover the source, addition time, and evaluation time of each statement added to a knowledge item. For example, in \textit{Wikipedia},
there is an edit history for each \textit{Wikipedia} article; on Q\&A sites, all answers to a question are recorded; and, in \textit{Github}, there is a version control mechanism to keep track
of all changes.

\xhdr{Generative process for knowledge evolution}
Our hypothesis is that the temporal information related to statement additions and evaluations reflects both the reliability of knowledge items and the trustworthiness of
information sources. More specifically, our intuition is as follows:
\begin{itemize}
\denselist
\item[I.] A reliable knowledge item should be stable in the sense that new statement addition will be rare, and it is less likely to be changed compared to unreliable items. Such
notion of reliability should be reflected in the statement addition process---as a knowledge item becomes more reliable, the number of statement addition events within a unit of
time should be smaller.

\item[II.] A trustworthy information source should result in statements which are verified quickly and refuted rarely.
Such notion of trustworthiness should be therefore reflected in its statement evaluation time---the more trustworthy an information source is, the shorter (lon\-ger) the time it will take to verify (refute) its statements.
\end{itemize}

In our temporal point process modeling framework, we build on the above intuition to account for both information reliability and source trustworthiness.
In particular, for each knowledge item, we model the statement addition times  $\{t_i\}$ as a counting process whose intensity directly relates to the reliability
of the item---as a knowledge item becomes more reliable, it is less likely to be changed. Moreover, each addi\-tion time $t_i$ is marked by its information source
$s_i$ and its evaluation time $\tau_i$, which in turn depends on the source trustworthiness and also have an impact on the overall reliability of the knowledge
item---the verification (refutation) of statements supported by trustworthy sources result in an increase (decrease) of the  reliability of the knowledge item.

More in detail, for each knowledge item $d$, we represent the statement addition times $\{t_i\}$ as a counting process $N_d(t)$, which
counts the number of statements that have been added up to but not including time $t$. Thus, we characterize the statement addition process using its corresponding
intensity $\lambda_d^{*}(t)$ as
\begin{equation}
 \EE[dN_d(t) | \Hcal_d(t)]= \lambda_d^{*}(t)dt,
\end{equation}
which captures the evolution of the reliability of the know\-ledge item over time. Here, the smaller the intensity $\lambda^*(t)$, the more reliable the knowledge
item at time $t$.
Moreover, since a knowledge item consists of a collection of statements, the overall reliability of the knowledge item will also depend on the individual reliability of its added statements
through their evaluations---the verification (refutation) of statements may result in an increase (decrease) of the reliability of the knowledge item, leading to an inhibition (increase) in the intensity of the statement additions to the
knowledge item.

Additionally, every time a statement $i$ is added to the knowledge item $d$, the corresponding information source  $s_i \in \Scal$ is sampled from a distribution $p(s | d)$ and
the evaluation time $\tau_i$ is sampled from a survival process, which we represent as a binary counting process $N_{i}(t) \in \{0,1\}$, in which $t=0$ corresponds to the time in which the statement is added and becomes one when $t=\tau_i-t_i$.
Here, we characterize this survival process using its corresponding intensity $\mu^{*}_i(t)$ as
%
\begin{equation}
 \EE[N_{i}(t) | \Hcal_d(t)] = \mu^{*}_i(t) dt,
\end{equation}
which captures the temporal evolution of the reliability of the $i$-th statement added to the knowledge item. Here, the smaller the intensity $\mu_i^*(t)$,
the shorter (longer) time it will take to verify (refute) it.
This intensity will depend, on the one hand, on the current intrinsic reliability of the corresponding knowledge item and, on the other hand, on the trustworthiness of the
source supporting the statement.

Next, we formally define the functional form of the intensities $\lambda_d^{*}(t)$ and $\mu_i^{*}(t)$, and the source distribution $p(s|d)$.

\xhdr{Knowledge item reliability}
For each knowledge item $d$, we consider the following form for its reliability function, or equivalently, its statement addition intensity:
\begin{align}\label{eq:intensity-addition}
\lambda_d (t) = \underbrace{\sum_j \phi_{d, j}  k(t-t_j)}_{\text{item intrinsic reliability}}+  \underbrace{\sum_{e_i \in \Hcal_d(t)}  \mathbf{w}^
\top_d \gammab_{s_i} g(t-\tau_i)}_{\text{effect of past evaluations}}.
\end{align}
In the above expression, the first term is a mixture of kernels $k(t)$ accounting for the temporal evolution of the intrinsic reliability of a knowledge item over time,
and the second term accounts for the effect that previous statement evaluations have on the overall reliability of the knowledge item.
Here, $\mathbf{w}_{d}$ and $\gammab_{s_i}$ are $L$-length vectors whose elements indicate, respectively, the weight (presence) of each topic in the know\-ledge item
and the per-topic influence of past evaluations of statements backed by source $s_i$. Finally, the function $g(t)$ is a nonnegative triggering kernel, which models the decay
of the influence of past evaluations over time.
If the evaluation is a refutation then we assume $\gammab_{s_i} \geq 0$, since a refuted statement typically decreases the reliability of the
knowledge item and thus triggers the arrival of new statements to replace it.
If the evaluation is a verification, we assume $\gammab_{s_i} \leq 0$, since a verified statement typically increases the reliability of the knowledge
item and thus inhibits the arrival of new statements to the knowledge item.
As a consequence, the above design results in an ``evaluation aware" process, which captures the effect that previous statement evaluations exert
on the reliability of a knowledge item.

\xhdr{Statement reliability}
As discussed above, every statement addition event $e_i$ is \emph{marked} with an evaluation time $\tau_i$, which we model using a survival process. The process
is ``statement driven" since it starts at the time when the statement addition event occurs and, within the process, $t=0$ corresponds to the addition time of the statement.
For each statement $i$, we adopt the following form for the statement reliability or, equivalently, for the intensity associated with its survival process:
\begin{align}\label{eq:lambda_survival}
\mu_i(t) =  {(1-N_i(t)) \Big[ \underbrace{\sum_j \beta_{d, j}  k(t+t_i-t_j)}_{\text{item intrinsic reliability}}+ \underbrace{\mathbf{w}^\top_{d} \alphab_{s_i}}_{\text{\stackanchor{source}{trustworthiness}}} \Big]}.
\end{align}
In the above expression, the first term is a mixture of kernels $k(t)$ accounting for the temporal evolution of the intrinsic reliability of the corresponding knowledge item $d$ and the
second term captures the trustworthiness of the source that supports the statement.
Here, $\mathbf{w}_{d}$ and $\alphab_{s_i}$ are L-length nonnegative vectors whose elements indicate, respectively, the weight (presence) of each topic in the knowledge
item $d$ and the trustworthiness of source $s_i$ in each topic.
Since the elements $\mathbf{w}_{d}$ sum up to one, the product  $\mathbf{w}_{d} \alphab_{s_i}$ can be seen as the average trustworthiness of the source $s_i$
in the knowledge item $d$.
With this modeling choice, the higher the parameter $ \alphab_{s_i}$, the quicker the evaluation of the statement.
Then, if the evaluation is a refutation, a high value of $\alphab_{s_i}$ implies low trustworthiness of the source $s_i$. In contrast, if it is a verification, a high value of $\alphab_{s_i}$ implies high trustworthiness.

Finally, note that the reliability of a statement, as defined in Eq.~\ref{eq:lambda_survival}, reflects how quickly (slowly) it will be refuted or verified,
and the reliability of a knowledge item, as defined in Eq.~\ref{eq:intensity-addition}, reflects how quickly (slowly) new statements are added to
the knowledge item.

\xhdr{Selection of source}
The source popularity $p(s|d)$ typically depends on the topics contained in the knowledge item $d$. Therefore, we consider the following form for the source distribution:
\begin{equation} \label{eq:source_dist}
p(s|d)= \sum_{\ell=1}^L w_{d,\ell} p(s|\ell), 
\end{equation}
where $w_{d,\ell} $ denotes the weight of topic $\ell$ in knowledge item $d$ and $p(s|\ell) \propto Multinomial(\pib_\ell)$ is the distribution of the sources for topic $\ell$, \ie, the vector $\pib_\ell$ contains the probability of each source to be assigned to a topic $\ell$.

\section{Efficient Parameter Estimation}
\label{sec:estimation}
In this section, we show how to efficiently learn the pa\-ra\-me\-ters of our model, as defined by Eqs.~\ref{eq:intensity-addition} and~\ref{eq:lambda_survival}, from a
set of statement addition and evaluation events.
%
Here, we assume that the topic weight vectors $\mathbf{w}_d$ are given\footnote{\scriptsize There are many topic modeling tools to learn the topic weight vectors
$\mathbf{w}_d$.}.
More specifically, given a set of sources $\Scal$ and a set of knowledge items $\Dcal$ with histories $\{\Hcal_1(T), \ldots, \Hcal_{|\Dcal|}(T)\}$, spanning a time period $[0,T)$, we find the
model parameters $\{\pib_\ell \}_{\ell=1}^L$, $\{\betab_d\}_{d=1}^{|\Dcal |}$, $\{\phib_d\}_{d=1}^{|\Dcal|}$, $\{\alphab_s\}_{s=1}^{|\Scal |}$ and $\{\gammab_s\}_{s=1}^{|\Scal |}$,
by solving the following maximum likelihood estimation (MLE) problem
\begin{align} \label{eqn:MLEopt}
        {\text{maximize }} & \Lcal(\{\pib_\ell \}, \{\betab_d\}, \{\phib_d\}, \{\alphab_s\}, \{\gammab_s\}) \nonumber \\
        {\text{subject to }} & \pib_{\ell} \geqslant 0,\, \betab_{d} \geqslant 0,\, \phib_{d} \geqslant 0,  \alphab_{s} \geqslant 0,\, \nonumber \mathbf{1}^{T} \pib_{\ell} = 1 \nonumber
\end{align}
where the log-likelihood is given by
\begin{equation}
\begin{split}
 &\Lcal =
  \sum_{d=1}^{|\Dcal | } \sum_{i: e_i \in \Hcal_d(T)} \log \underbrace{p(t_i |  \Hcal_d(t_i), \phib_d, \{\gammab_s\}, \mathbf{w}_d)}_{\text{statements additions}}
  \quad + \sum_{d=1}^{|\Dcal | } \sum_{i: e_i \in \Hcal_d(T)} \log \underbrace{p(\Delta_i | t_i, \betab_d, \{\alphab_s\}, \mathbf{w}_d)}_{\text{statements evaluations}}
   \quad \\
   &\quad+\sum_{d=1}^{|\Dcal | } \sum_{i: e_i \in \Hcal_d(T)} \log \underbrace{p(s_i | \{\pib_\ell \}, \mathbf{w}_d)}_{\text{sources popularity}}.
\end{split}
\end{equation}
In the above likelihood, the first term accounts for the times at which statements are added to the knowledge item,
the second term accounts for the times at which statements are evaluated,
and the third term accounts for the probability that source $s_i$ is assigned to the statement addition event $e_i$.
Since the first two terms correspond to likelihoods of temporal point processes, they can be computed using Eq.~\ref{eq:loglikehood_fun}. The third term is
simply given by $p(s_i | \{\pib_\ell \}_{\ell=1}^L, \mathbf{w}_d) = \sum_{\ell=1}^L w_{d,\ell} \pib_{\ell}(s_i)$, where $\pib_{\ell} (s_i)$ denotes the $s_i$-th element
of $\pib_\ell$.

Remarkably, the above terms can be expressed as linear combinations of logarithms and linear functions or compositions of linear functions with logarithms
and thus easily follow that the above optimization problem is jointly convex in all the parameters.
Moreover, the problem can be decomposed into three independent problems, which can be solved in parallel obtaining local solutions that are in turn globally
optimal.
For knowledge repositories using refutation, \ie, $\gammab_{s} \geqslant 0$, we solve both the first and second problem by adapting the algorithm by Zhou
et al.~\cite{zhou2013learning}.
For knowledge repositories using verification, \ie, $\gammab_{s} \leqslant 0$, we solve the first problem using cvxpy~\cite{cvxpy} and the second problem
by adapting the algorithm by Zhou et al. ~\cite{zhou2013learning}.
In both cases, the third problem can be computed analytically as
\begin{equation} \label{eqn:MLEpi}
\pib_{\ell} (s)= \frac{\sum_{d=1}^{|\Dcal | } \mathbf{w}_d (\ell) \hat{\pib}_d(s)}{\sum_{d=1}^{|\Dcal | } \sum_{s'=1}^{|\Scal |} \mathbf{w}_d (\ell) \hat{\pib}_d(s')},
\end{equation}
where $\mathbf{w}_d (\ell)$ denotes the $\ell$-th element of $\mathbf{w}_d$, and $\hat{\pib}_d(s)$ is the probability that source $s$ is assigned to a statement in
knowledge item $d$. In particular, $\hat{\pib}_d(s)$ can be computed as
\begin{equation}
\hat{\pib}_d(s)= \frac{n_{d,s}}{\sum_{s'=1}^{|\Scal |} n_{d,s'}},
\end{equation}
where $n_{d,s}$ is the number of statement addition events in the history of the knowledge item that are backed by source $s$, \ie, $|\{e_i \in \Hcal_d(T)| s_i=s\}|$.
%
In practice, we found that adding a $\ell$-1 penalty term on the parameters $\{ \betab_d \}$, \ie, $\eta \sum_{d} ||\betab_d||_1$, which we set by cross-validation, avoids
overfitting and improves the predictive performance of our model.
\begin{figure}[t]
\centering
\subfloat[$\alpha$ ]{\makebox[0.23\textwidth][c]{\includegraphics[width=0.23\textwidth]{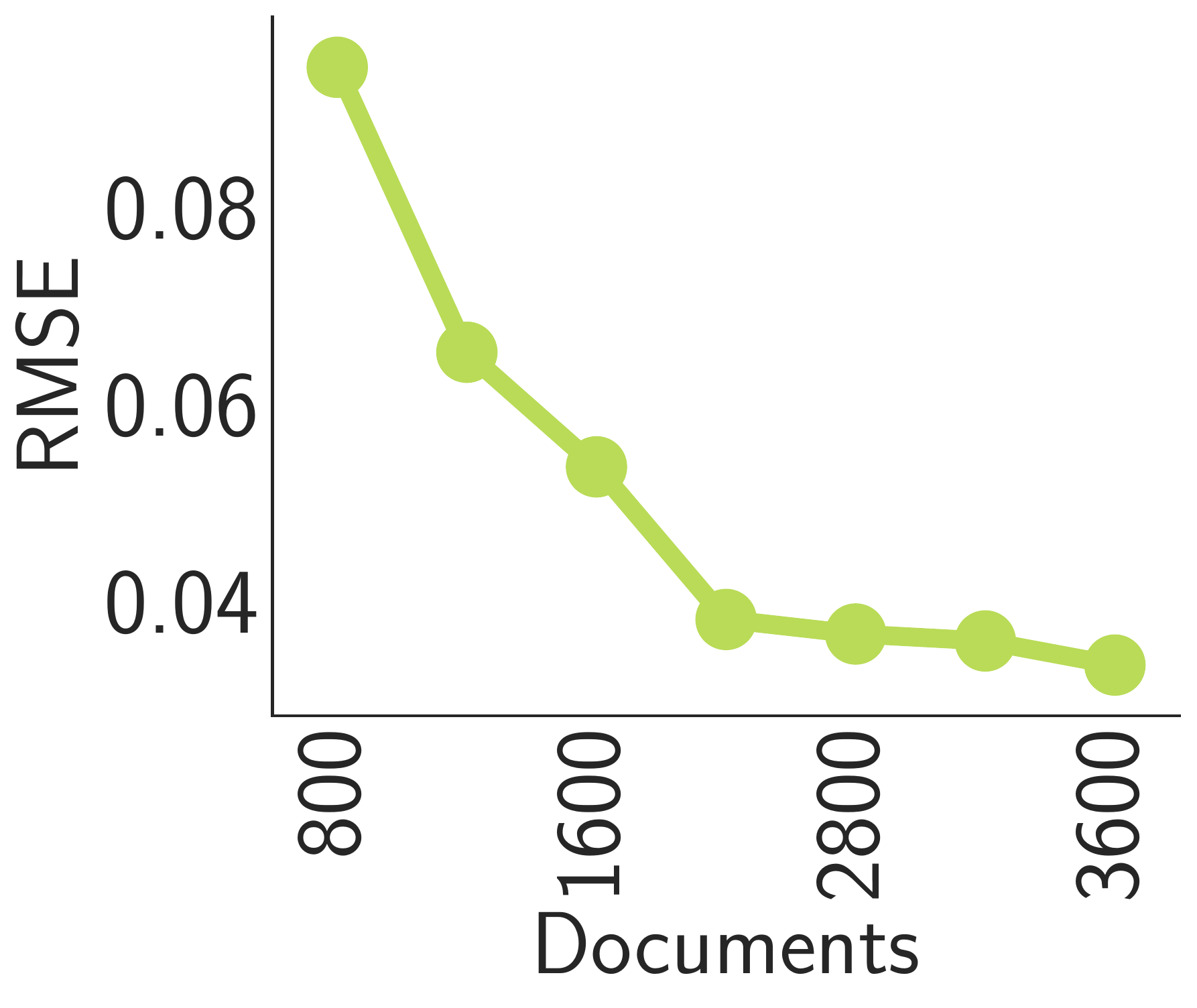}}\label{fig:alpha}}
\subfloat[$\gamma$ ]{\makebox[0.23\textwidth][c]{\includegraphics[width=0.23\textwidth]{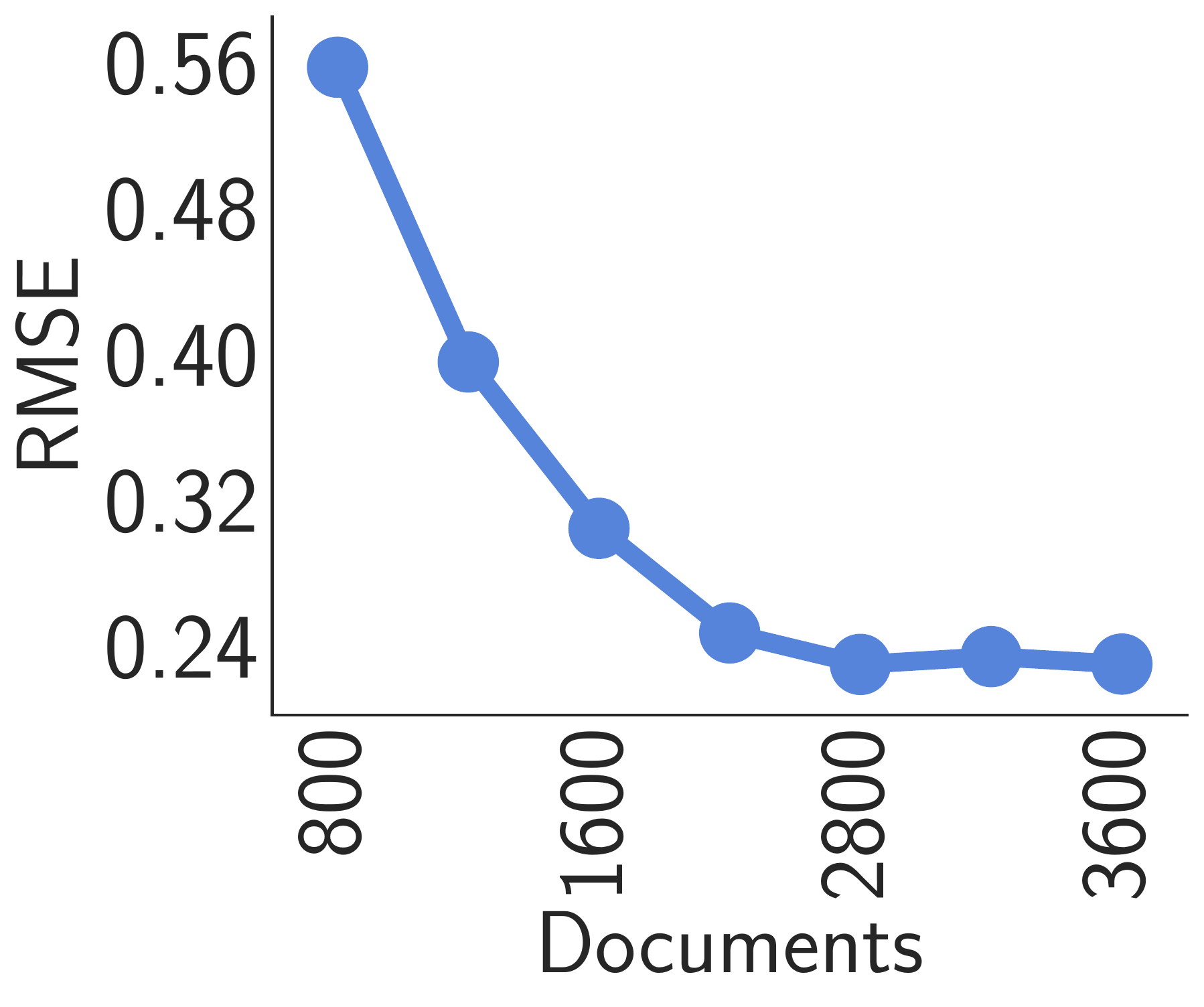}}\label{fig:gamma}}
\subfloat[$\beta$ ]{\makebox[0.23\textwidth][c]{\includegraphics[width=0.23\textwidth]{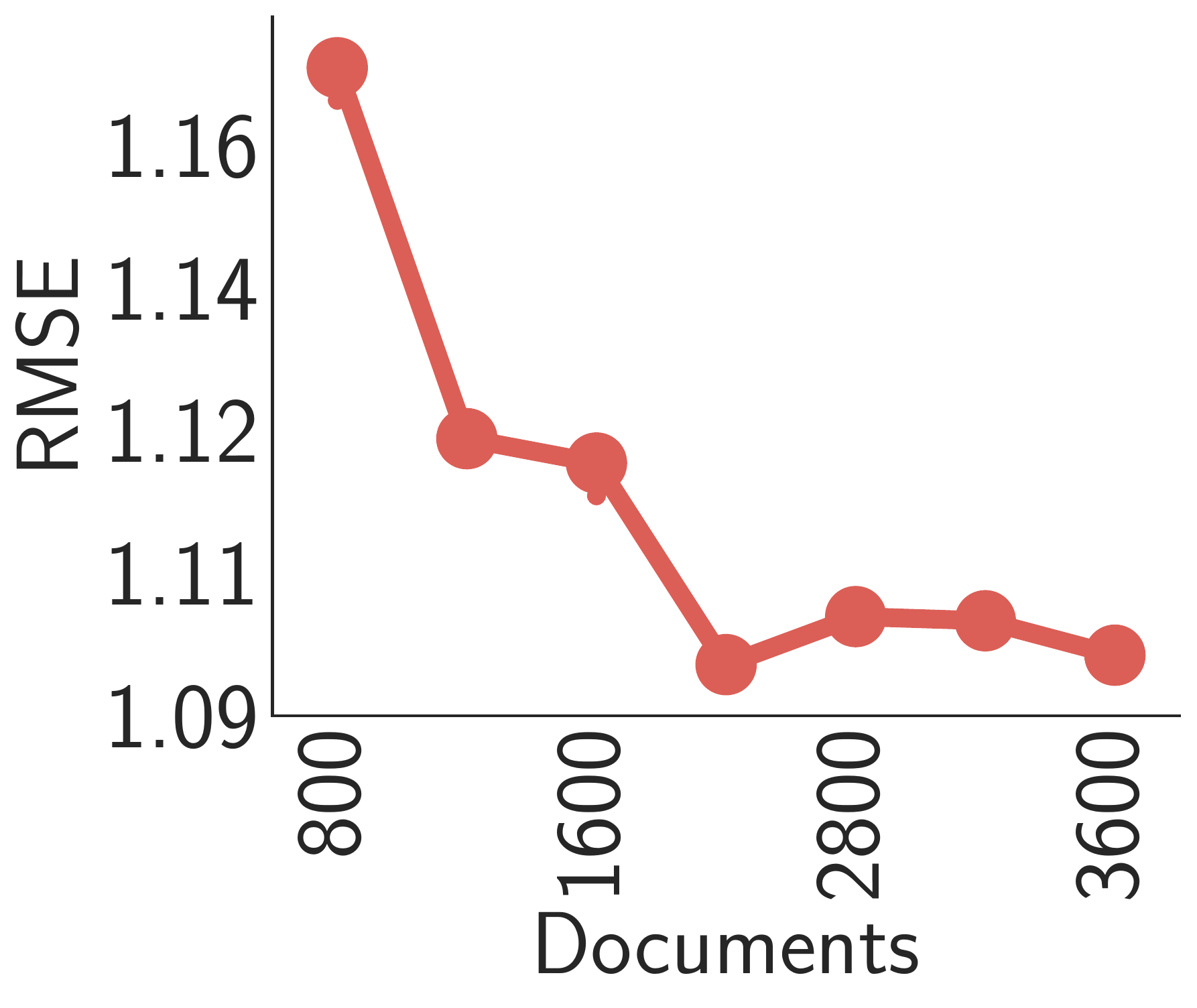}}\label{fig:beta}}
\subfloat[$\phi$]{\makebox[0.23\textwidth][c]{\includegraphics[width=0.23\textwidth]{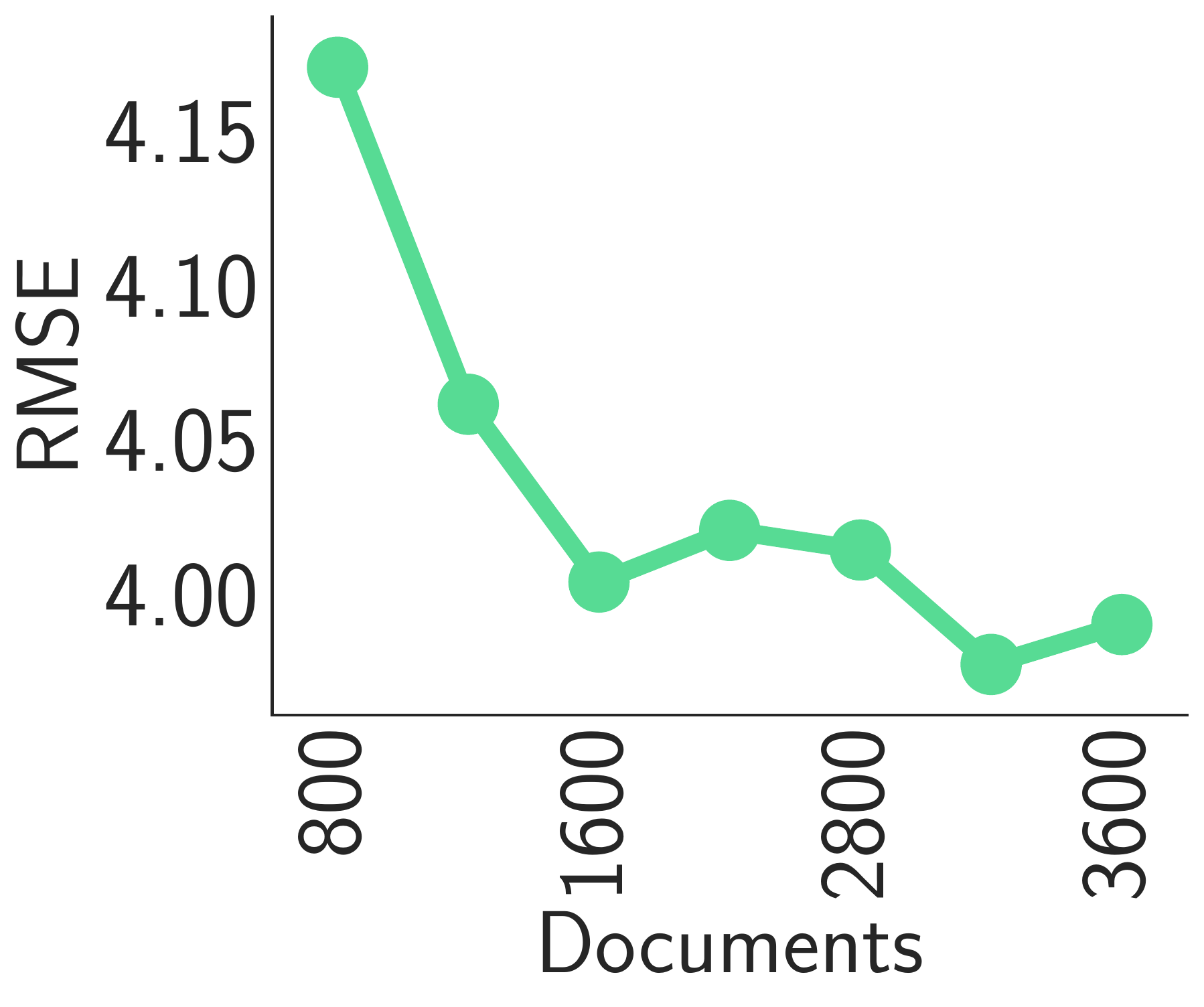}}\label{fig:phi}}
 \caption{Performance of our model parameter estimation method on synthetic data in terms of root mean squared error (RMSE). The estimation
 becomes more accurate as we feed more knowledge items into our estimation procedure. However, since each new knowledge item increases the number
 of $\beta$ and $\phi$ parameters, once the source parameter estimation becomes accurate enough, the estimation error for $\beta$ and $\phi$
 flattens.}
        \label{fig:rellerr-vs-documents}
\end{figure}

\section{Experiments on Synthetic Data}
\label{sec:experiments-synthetic}
Our goal in this section is to investigate if our parameter estimation method can accurately recover the true model parameters from statement addition and evaluation events.
We examine this question using a synthetically generated dataset from our probabilistic model.

\xhdr{Experimental setup}
We set the number of sources to $|\Scal|=400$, the total number of knowledge items to $|\Dcal| = 3{,}600$, and assume the evaluation mechanism is refutation. We assume
there is only one topic and then, for each source, we sample its trustworthiness $\alpha_s$ from the Beta distribution $Beta(2.0, 5.0)$ and its parameter $\gamma_s$ from
the uniform distribution $U(0, b)$, where $b=0.03 \times \max(\{\alpha_s\}_{s\in \Scal})$.
For the temporal evolution of the intrinsic reliability in the addition and evaluation processes, we consider a mixture of three radial basis (RBF) kernels located at times
$t_j = 0, 6, 12$, with standard deviations of $2$ and $0.5$, respectively.
Then, for each knowledge item, we first pick one of the kernel locations $j$ uniformly at random, which determines the only \emph{active} kernel for both the addition and
the evaluation processes in the knowledge item, and sample their associated parameters, $\phi_{d,j}$ and $\beta_{d,j}$, from the log-normal distribution $\ln\mathcal{N}(3.5, 0.1)$
and the uniform distribution $U(0, 0.2\phi_d)$, respectively.
Moreover, we assume that only up to five (different) sources are active in each knowledge item, which we pick at random, and then draw a source probability vector for these five
active sources in the knowledge item from a Dirichlet distribution with parameter $0.5$.
The choice of prior distributions for the model parameters ensures enough variability across knowledge items and sources, so that the model parameters can be recovered.
Finally, we generate addition and refutation samples from the resulting addition and evaluation processes during the time interval $(0,15]$.

\xhdr{Results}
We evaluate the accuracy of our model estimation procedure by means of the root mean square error (RMSE) between the true $(x)$ and the estimated $(\hat{x})$ parameters, \ie, $\text{RMSE}(x)= \sqrt{\EE[(x-\hat{x})^2] }$.
Figure~\ref{fig:rellerr-vs-documents} shows the parameter estimation error with respect to the number of knowledge items used to estimate the model parameters.
Since the source parameters $\alpha$ and $\gamma $ are shared across knowledge items, the estimation becomes more accurate as we feed more knowledge items
into our estimation procedure.
However, every time we observe a new knowledge item, the number of parameters increases with an additional $\beta_d$ and $\phi_d$. Therefore, the knowledge item
parameter estimation only becomes more accurate as a consequence of a better estimation of the source parameters. As soon as the source parameter
estimation becomes \emph{good enough}, the estimation does not improve further and the estimation error flattens.
\begin{figure}[t]
  \centering
       \begin{tabular}{c c}
       \includegraphics[width=0.30\textwidth]{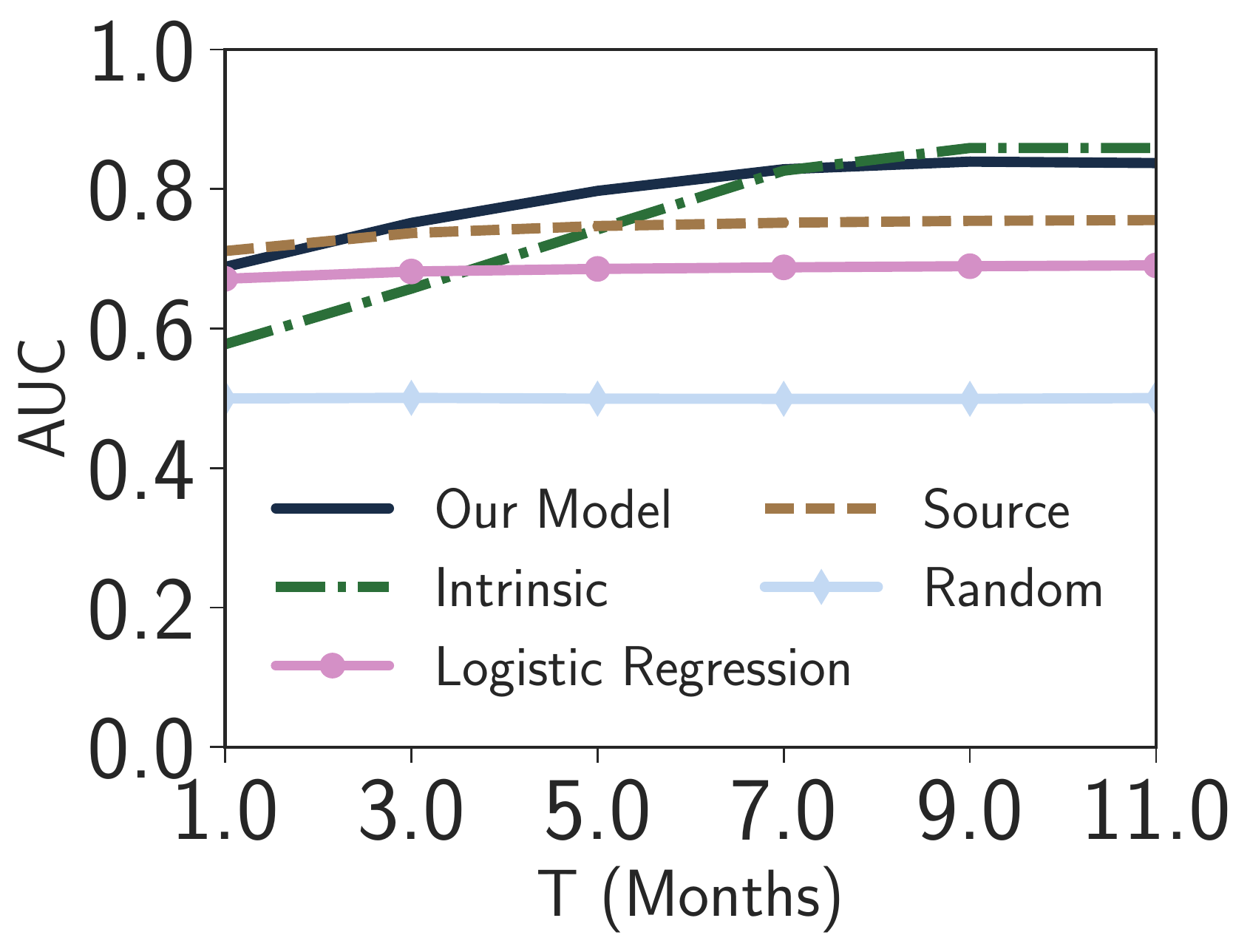} \hspace{17mm} &
       \includegraphics[width=0.30\textwidth]{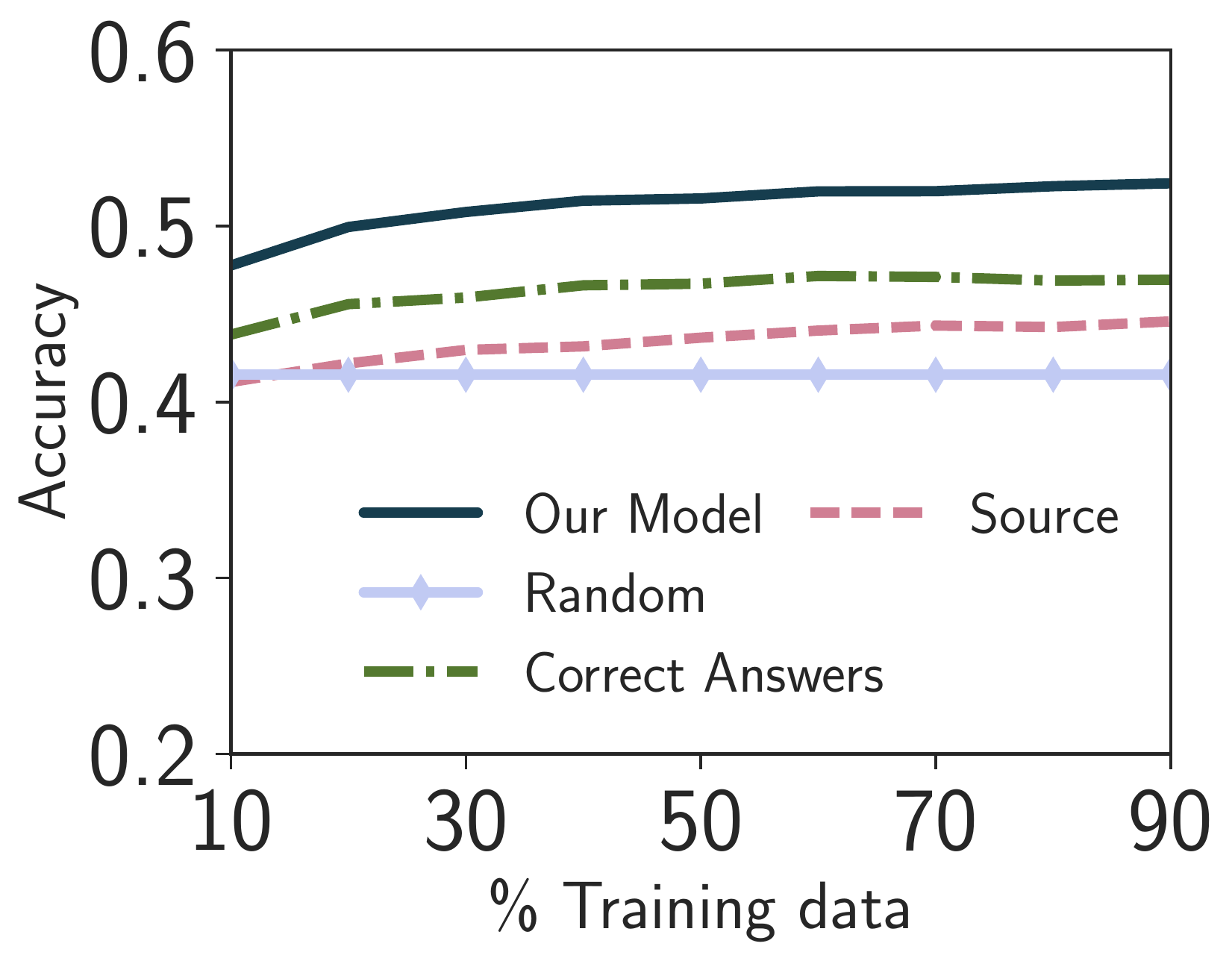}\\
        (a) \textit{Wikipedia} & (b) \textit{Stack Overflow}
        \end{tabular}
       \caption{Prediction performance.
       Panel (a) shows the AUC achieved by our model and three baselines (Intrinsic, Source and Logistic Regression) for predicting whether a statement will be removed (refuted) from a \textit{Wikipedia} article within a time period of $T$ after it is posted; for different values of $T$.
       Panel (b) shows the success probability achieved by our model and two baseline (Source and Correct Answer) at predicting which answer to a question, among several answers, will be eventually
       verified in \textit{Stack Overflow}.}
       \label{fig:rocauc}
\end{figure}

\section{Experiments on Real Data}
\label{sec:experiments-real}
In this section, we apply our model estimation method to large-scale data gathered from two knowledge repositories: \textit{Wikipedia}, which uses refutation as evaluation
mechanism (\ie, deleted statements), and \textit{Stack Overflow}, which uses verification (\ie, accepted answers).
First, we show that our model can accurately predict whether a particular statement in a \textit{Wikipedia} article will be refuted after a certain
period of time, as well as which of the answers to a question in \textit{Stack Overflow} will be accepted.
Then, we show that it provides meaningful measures of web source trustworthiness in \textit{Wikipedia} and user trustworthiness in \textit{Stack Overflow}.
Finally, we demonstrate that our model can be used to: (i) pinpoint the changes on the intrinsic reliability of a \textit{Wikipedia} article over time and these changes match
external noteworthy controversial events; and, (ii) find questions and answers in \textit{Stack Overflow} with similar levels of intrinsic reliability, which in this case correspond
to popularity and difficulty.

\xhdr{Data description and methodology}
To build our \textit{Wi\-ki\-pe\-dia} dataset, we ga\-ther complete edit history, up to July 8, 2014, for $~1$ million \textit{Wikipedia} English articles and track all the references
(or links) to sources within each of the edits.
Then, for each article $d$, we record for each added statement, its associated source $s_i$, its addition time $t_i$, and its refutation (deletion) time $\tau_i$, if
any. Such recorded data allows us to reconstruct the history of each article (or knowledge item), as given by Eq.~\ref{eq:history}.
Moreover, since we can only expect our model estimation method to provide reliable and accurate results for articles and web sources with enough number of events,
we only consider articles with at least $20$ link additions and web sources that are used in at least $10$ references.
After these preprocessing steps, our dataset consists of $\sim$$50$ thousand web sources that appeared in $\sim$$100$ thousand articles, by means of $\sim$$10.4$
million addition events and $\sim$$9$ million refutation (deletion) events. The significant drop in the number of articles can be attributed to the large number of incomplete articles
on Wikipedia, which lack reasonable number of citations.
Finally, we run the (Python library) Gensim~\cite{gensim} on the la\-test revision of all documents in the dataset, with 10 topics and default
parameters, to obtain the topic weight vectors $\mathbf{w}_d$, and apply our model estimation method, described in Section~\ref{sec:estimation}.
Both in Eqs.~\ref{eq:intensity-addition} and~\ref{eq:lambda_survival}, we used $19$ RBF kernels, spaced every $9$ months with standard deviation of $3$ months.
In Eq.~\ref{eq:intensity-addition}, we used exponential triggering kernels with $\omega = 0.5$ hours$^{-1}$.

To build our \textit{Stack Overflow} dataset, we gathered history of answers from January 1, 2011 up to June 30, 2011, for $\sim$$500$ thousand questions \footnote{\scriptsize Dataset available at \url{https://archive.org/details/stackexchange}.}.
Then, for each answer, we record the question $d$ it belongs to, the user $s_i$ who posted the answer, its addition time $t_i$, and its verification (acceptance) time $\tau_i$,
if any.
Similarly as in the \textit{Wikipedia} dataset, such recorded data allows us to reconstruct the history of each question (or knowledge item), as given by Eq.~\ref{eq:history}.
Again, since our model estimation method can only provide reliable and accurate results for questions and users with enough number of events, we only consider
questions with an accepted answer (if any exist) within $4$ days of publication time and users who posted at least $4$ accepted answers.
After these preprocessing steps, our data consists of $\sim$$378$ thousand questions which accumulate $\sim$$724$ thousand addition events (answers) and $\sim$$224$ thousand verification events (accepted answers).
In this case, we assume a single topic and therefore the weight vector $\mathbf{w}_d$ becomes a scalar value of $1$.
Finally, we apply our model estimation method, described in Section~\ref{sec:estimation}.
In this case, in Eqs.~\ref{eq:intensity-addition} and~\ref{eq:lambda_survival}, we used single constant kernels $\beta_{d}$ and $\phi_{d}$, respectively,
since the intrinsic reliability of questions in \textit{Stack Overflow} does not typically change over time.
In Eq.~\ref{eq:intensity-addition}, we used step functions as triggering kernels, since the inhibiting effect of an accepted answer does not decay over time.

In both datasets, our parameter estimation method runs in $\sim$$4$ hours using a single
 machine with 10 cores and 64 GB RAM.

\begin{table}[t]
  \centering
  \small
  \setlength\tabcolsep{3pt}
  \begin{tabular}{c c c c c c c c c c c}
    \toprule
    \multicolumn{7}{c|}{Music} & \multicolumn{4}{c}{Politics} \\
    \hhline{======|=====}
    \multicolumn{1}{c}{Rank}&\multicolumn{2}{c}{domain}&\multicolumn{3}{c|}{Pr. rm. in} & \multicolumn{2}{c}{domain}&\multicolumn{3}{c}{Pr rm. in} \\
    \multicolumn{1}{c}{}&\multicolumn{2}{c}{}&\multicolumn{3}{c|}{6 months} & \multicolumn{2}{c}{}&\multicolumn{3}{c}{6 months} \\
    \hhline{-----|------}
    \multicolumn{1}{c}{1} &
    \multicolumn{2}{c}{guardian.co.uk} &
    \multicolumn{3}{c|}{$0.15$} &

    \multicolumn{2}{c}{nytimes.com}&
    \multicolumn{3}{c}{$0.18$} \\

    \multicolumn{1}{c}{2} &
    \multicolumn{2}{c}{rollingstone.com} &
    \multicolumn{3}{c|}{$0.17$} &

    \multicolumn{2}{c}{guardian.co.uk}&
    \multicolumn{3}{c}{$0.19$} \\

    \multicolumn{1}{c}{3} &
    \multicolumn{2}{c}{nytimes.com} &
    \multicolumn{3}{c|}{$0.17$} &

    \multicolumn{2}{c}{google.com}&
    \multicolumn{3}{c}{$0.20$} \\

    \multicolumn{1}{c}{6} &
    \multicolumn{2}{c}{billboard.com} &
    \multicolumn{3}{c|}{$0.26$} &

    \multicolumn{2}{c}{usatoday.com}&
    \multicolumn{3}{c}{$0.24$} \\

    \multicolumn{1}{c}{13} &
    \multicolumn{2}{c}{mtv.com} &
    \multicolumn{3}{c|}{$0.32$} &

    \multicolumn{2}{c}{whitehouse.gov}&
    \multicolumn{3}{c}{$0.29$}  \\

    \multicolumn{1}{c}{Last} &
    \multicolumn{2}{c}{twitter.com} &
    \multicolumn{3}{c|}{$0.56$} &

    \multicolumn{2}{c}{cia.com}&
    \multicolumn{3}{c}{$0.45$} \\
  \bottomrule
  \end{tabular}
  \caption{Top 20 most popular web sources from \textit{Wikipedia} in each topic ranked by the probability that a link from them is removed within 6 months (Most reliable on top).} 
  \label{fig:unrel-wiki-table}
\end{table}

\xhdrq{Can we predict if a statement will be removed from Wikipedia?}
Our model can answer this question by solving a binary classification problem: predict whether a statement will be removed (refuted) within a
time period of $T$ after it is posted.

\emph{--- Experimental setup:}
We first split all addition events into a training set ($90$\% of the data) and a test set (the remaining $10$\%) at
random, then fit the parameters of the information survival processes given by Eq.~\ref{eq:lambda_survival} using
only the evaluation times of the addition events from the training set, and finally predict whether particular statements
in the test set will be removed  within a time period of $T$ after it is posted.
We  compare the performance of our model with three baselines: ``Intrinsic", ``Source" and ``Logistic Regression."
``Intrinsic" attributes all changes in an article to the intrinsic (un)reliability of that document. We can capture this assumption in our model by assuming that the parameter $\alphab_s$ in
Eq.~\ref{eq:lambda_survival} is set to zero.
Inspired by the model proposed by Adler and De Alfaro~\cite{adler2007content}, we implement the baseline ``Source'', which only accounts for the trustworthiness of the source that supports a statement, \ie, it assumes that the intrinsic reliability of the article, parametrized by $\betab_d$ in Eq.~\ref{eq:lambda_survival},
is set to zero.
Finally, ``Logistic Regression" is a logistic regression model that uses the source identity (in one-hot representation), the document topic vector and the addition time of links as features. Here, we train a different logistic regression model per time window.

\emph{--- Results:}
Since the dataset is highly unbalanced (only $25\%$ of statements in the test set survive longer than $6$ months), we evaluate the classification accuracy  in terms of the area under the ROC curve (AUC), a standard metric for quantifying
classification performance on unbalanced data.
Figure~\ref{fig:rocauc}(a) shows the AUC achieved by our model and the baselines for different values of $T$.
Our model always achieves AUC values over $0.69$, it improves its performance as $T$ increases, and outperforms all baselines across the full spectrum of values of $T$.
The ``Source" baseline exhibits a comparable performance to our method for low values of $T$, however, its performance barely improves as $T$ increases, in
contrast, the ``Intrinsic" baseline performs poorly for low values of $T$ but exhibits a comparable performance to our method for high values of $T$.
Finally, ``Logistic Regression" achieves an AUC lower than our method across the full spectrum of values of $T$.

The above results suggest that refutations that occur quick\-ly after a statement is posted, are mainly due to the untrustworthiness of the source; while refutations that
occur later in time are due to the intrinsic unreliability of the article.
As a consequence, our model, by accounting for both source trustworthiness and intrinsic reliability of information, can predict both quick and slow refutations more
accurately than models based only on one of these two factors.

\begin{table}[t]
  \centering
  \small
  \begin{tabular}{c c c c c c c}
    \toprule
    \multicolumn{7}{c}{Stack Overflow}\\
    \hhline{=======}
    \multicolumn{1}{c}{Rank}&\multicolumn{2}{c}{user-id}&\multicolumn{2}{c}{ranking}&\multicolumn{2}{c}{P accept} \\
    \multicolumn{1}{c}{}&\multicolumn{2}{c}{}&\multicolumn{2}{c}{}&\multicolumn{2}{c}{in 4 days} \\
    \hhline{-------}

    \multicolumn{1}{c}{1} &
    \multicolumn{2}{c}{318425}&
    \multicolumn{2}{c}{top 0.30\%}&
    \multicolumn{2}{c}{$0.93$} \\

    \multicolumn{1}{c}{2} &
    \multicolumn{2}{c}{405015}&
    \multicolumn{2}{c}{top 0.07\%}&
    \multicolumn{2}{c}{$0.81$} \\

    \multicolumn{1}{c}{3} &
    \multicolumn{2}{c}{224671}&
    \multicolumn{2}{c}{top 0.01\%}&
    \multicolumn{2}{c}{$0.81$} \\

    \multicolumn{1}{c}{138} &
    \multicolumn{2}{c}{246342}&
    \multicolumn{2}{c}{top 0.12\%}&
    \multicolumn{2}{c}{$0.53$} \\

    \multicolumn{1}{c}{139} &
    \multicolumn{2}{c}{616700}&
    \multicolumn{2}{c}{top 0.36\%}&
    \multicolumn{2}{c}{$0.53$} \\

    \multicolumn{1}{c}{Last} &
    \multicolumn{2}{c}{344491}&
    \multicolumn{2}{c}{top 0.97\%}&
    \multicolumn{2}{c}{$0.53$} \\
  \bottomrule
  \end{tabular}
  \caption{\textit{Stack Overflow} users with more than $100$ answers (140 users) ranked by the probability that answer they provide is verified within 4 days (Most reliable on top).
  The table also shows the ranking provided by \textit{Stack Overflow}.}
  \label{fig:unrel-so-table}
\end{table}

\begin{figure*}[t]
  \centering
   \subfloat[Analysis of parameter $\alphab$]{\setlength\tabcolsep{3pt}
  \begin{tabular}{c c c }
  \includegraphics[width=0.15\textwidth]{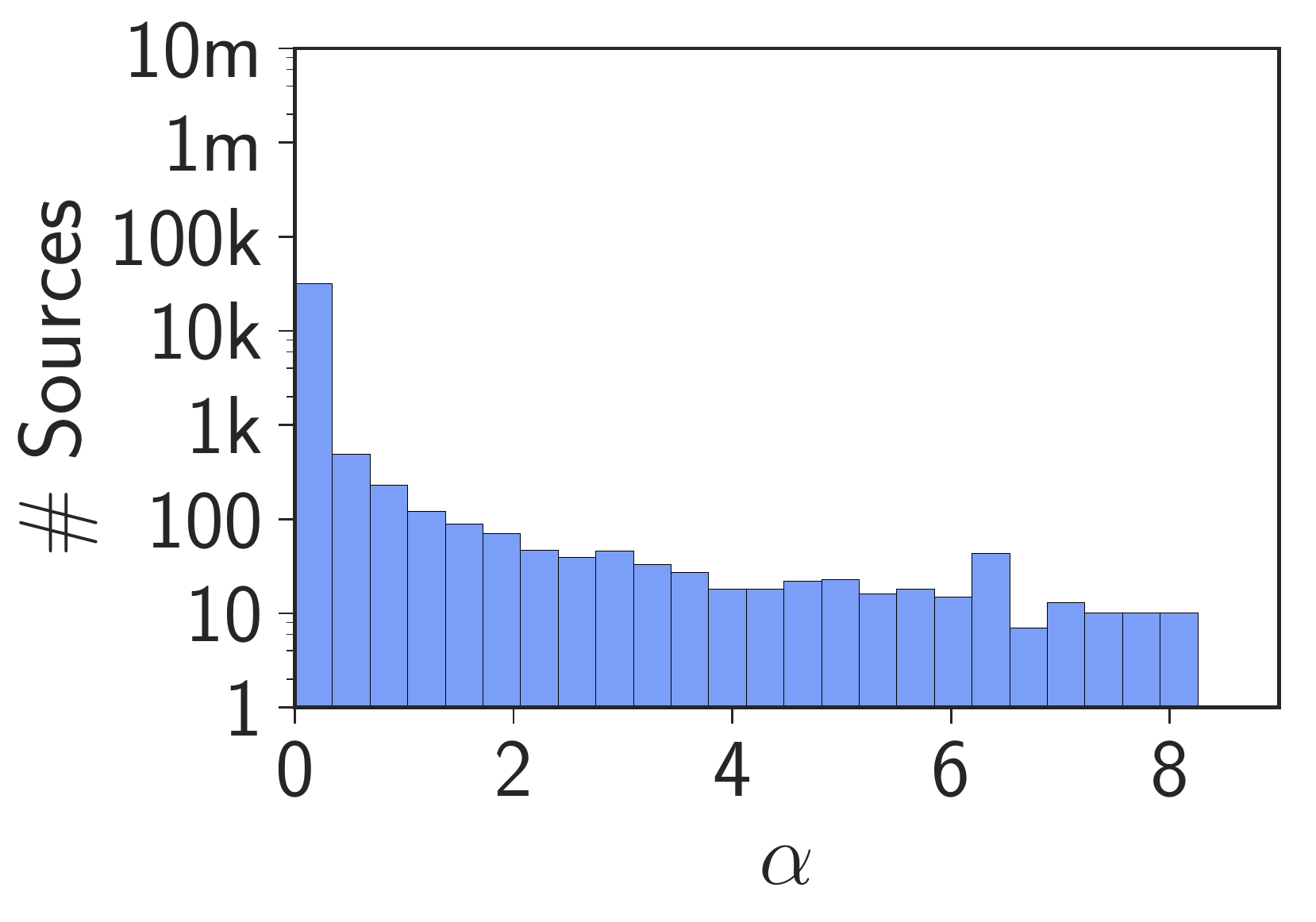} &
  \includegraphics[width=0.15\textwidth]{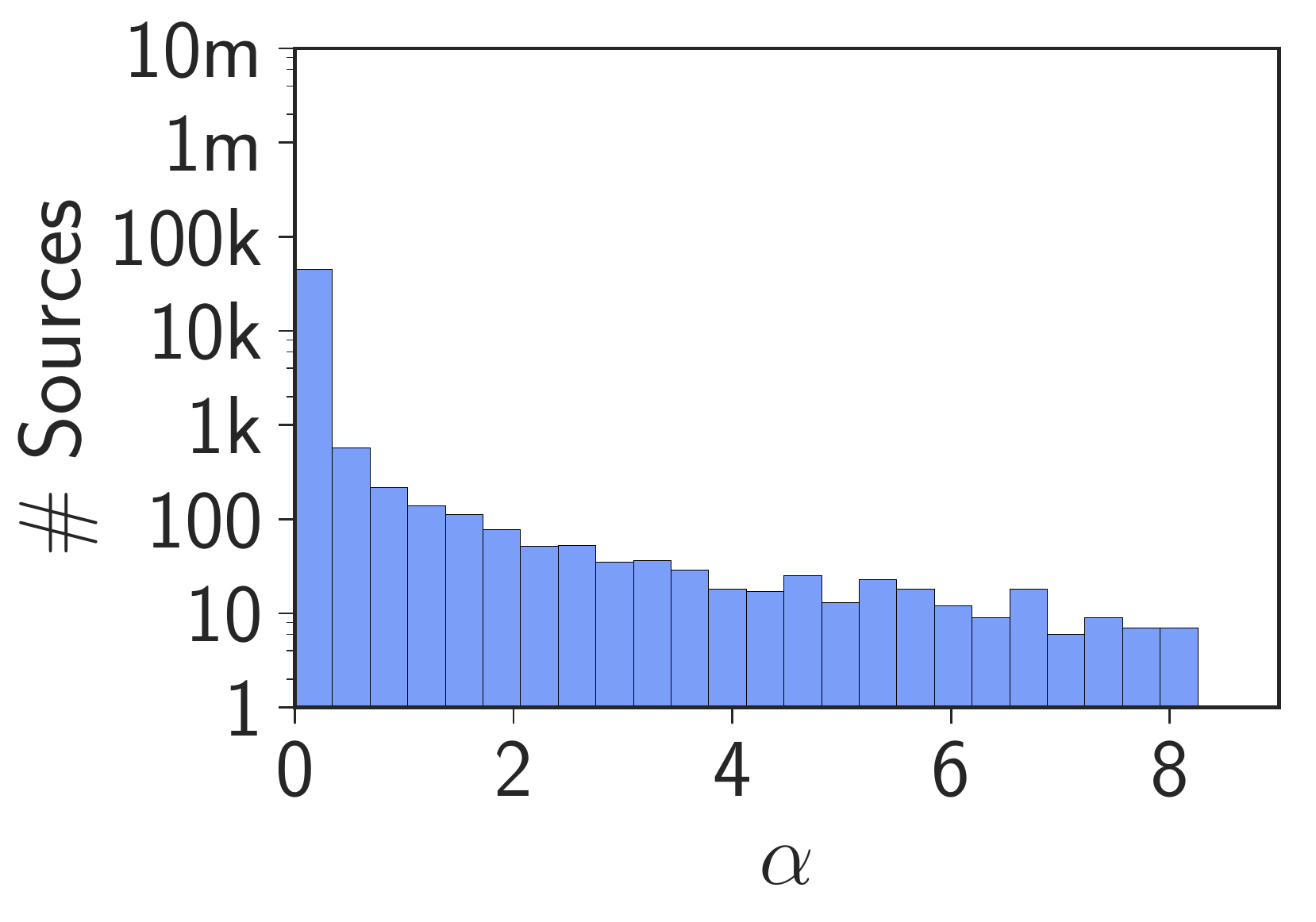} &
  \includegraphics[width=0.15\textwidth]{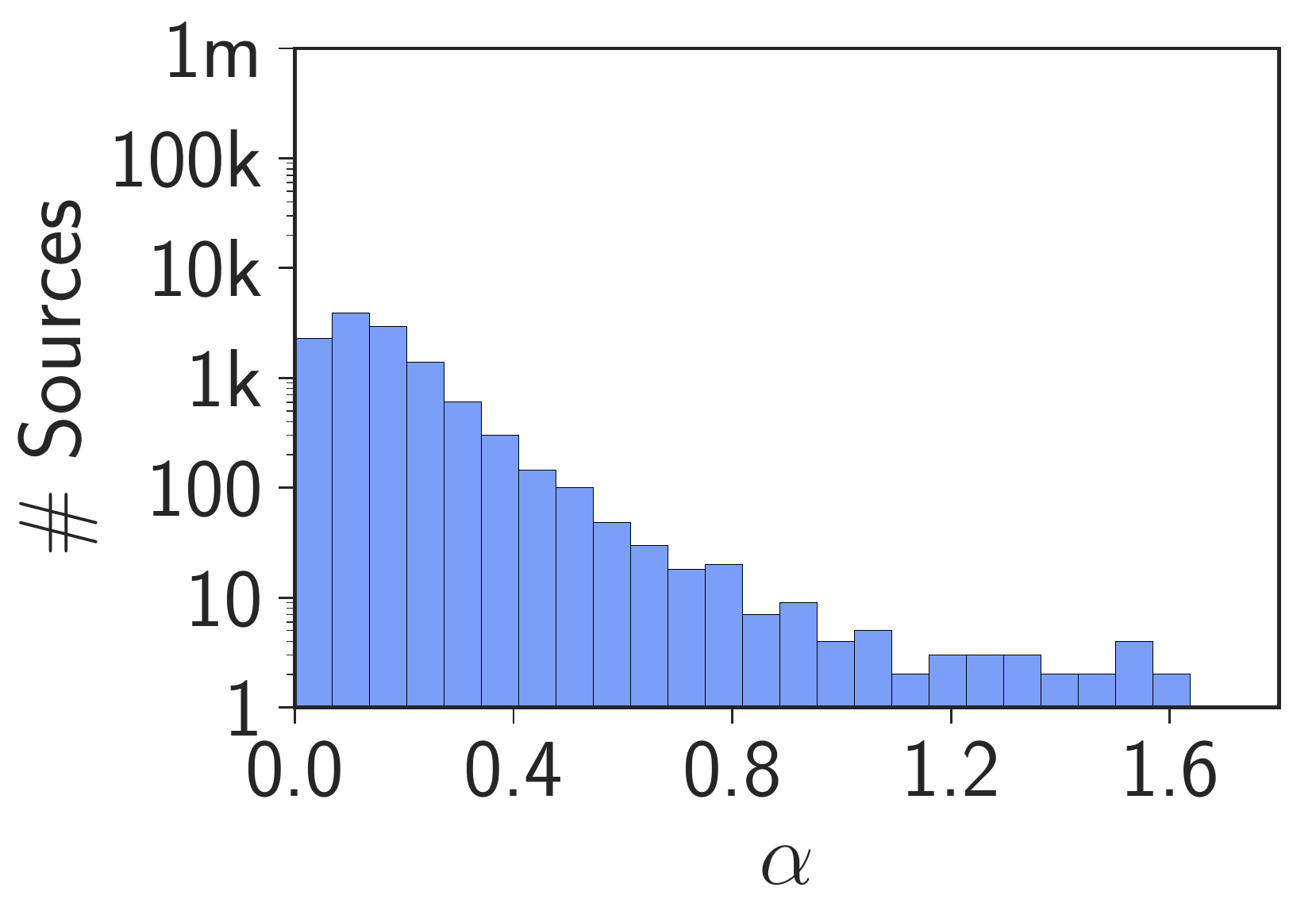} \\
  \includegraphics[width=0.15\textwidth]{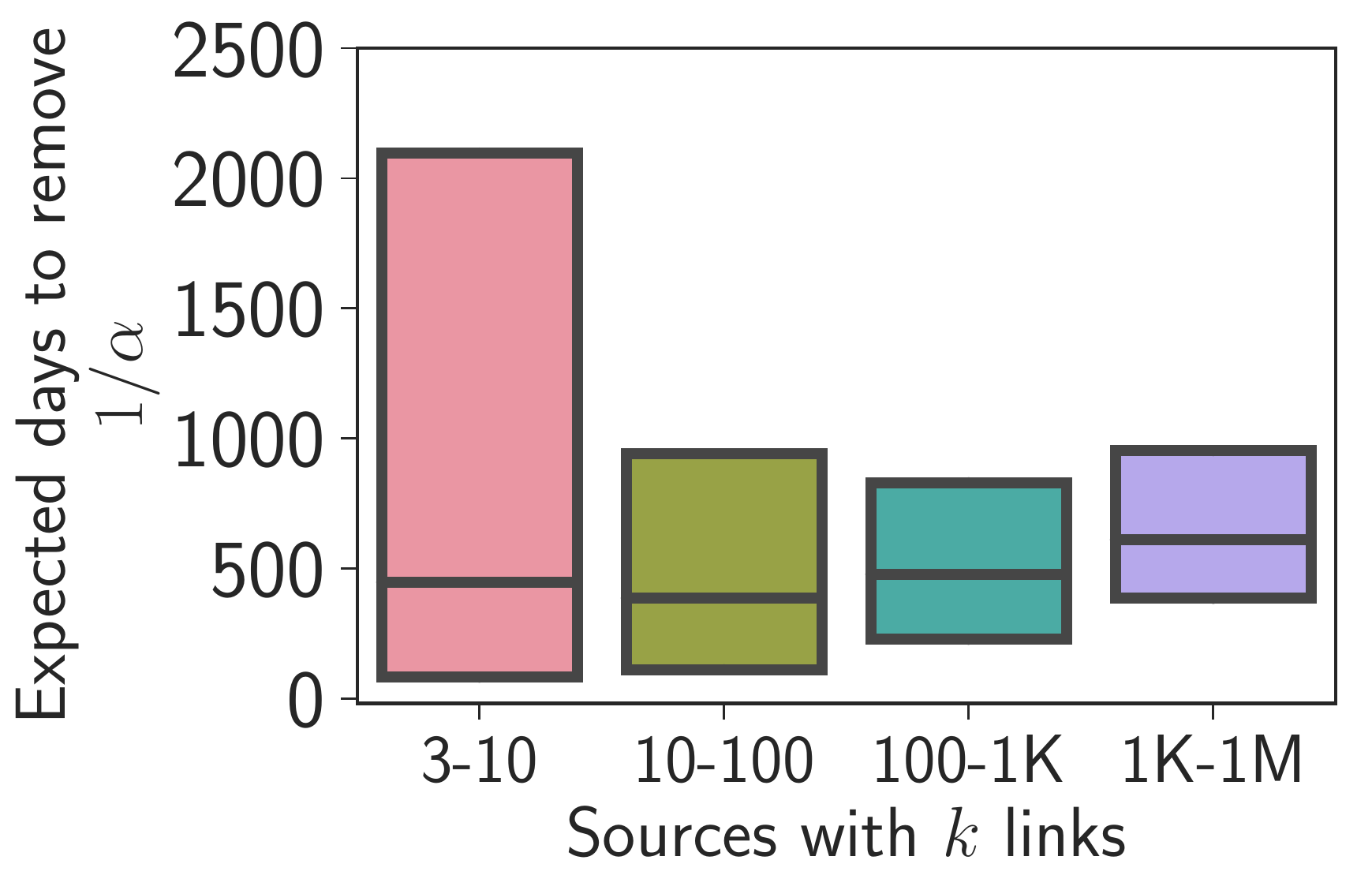} &
  \includegraphics[width=0.15\textwidth]{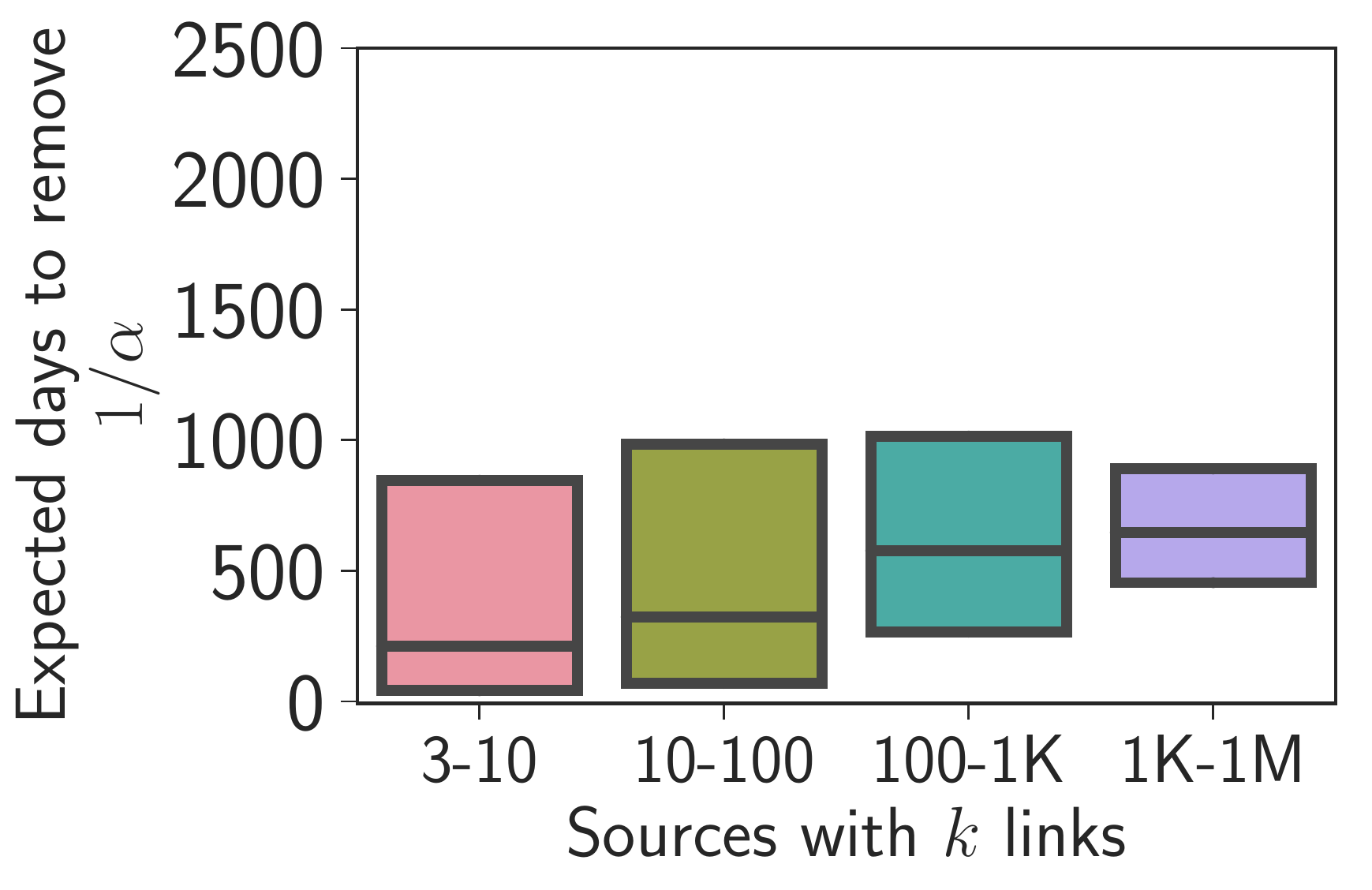} &
  \includegraphics[width=0.15\textwidth]{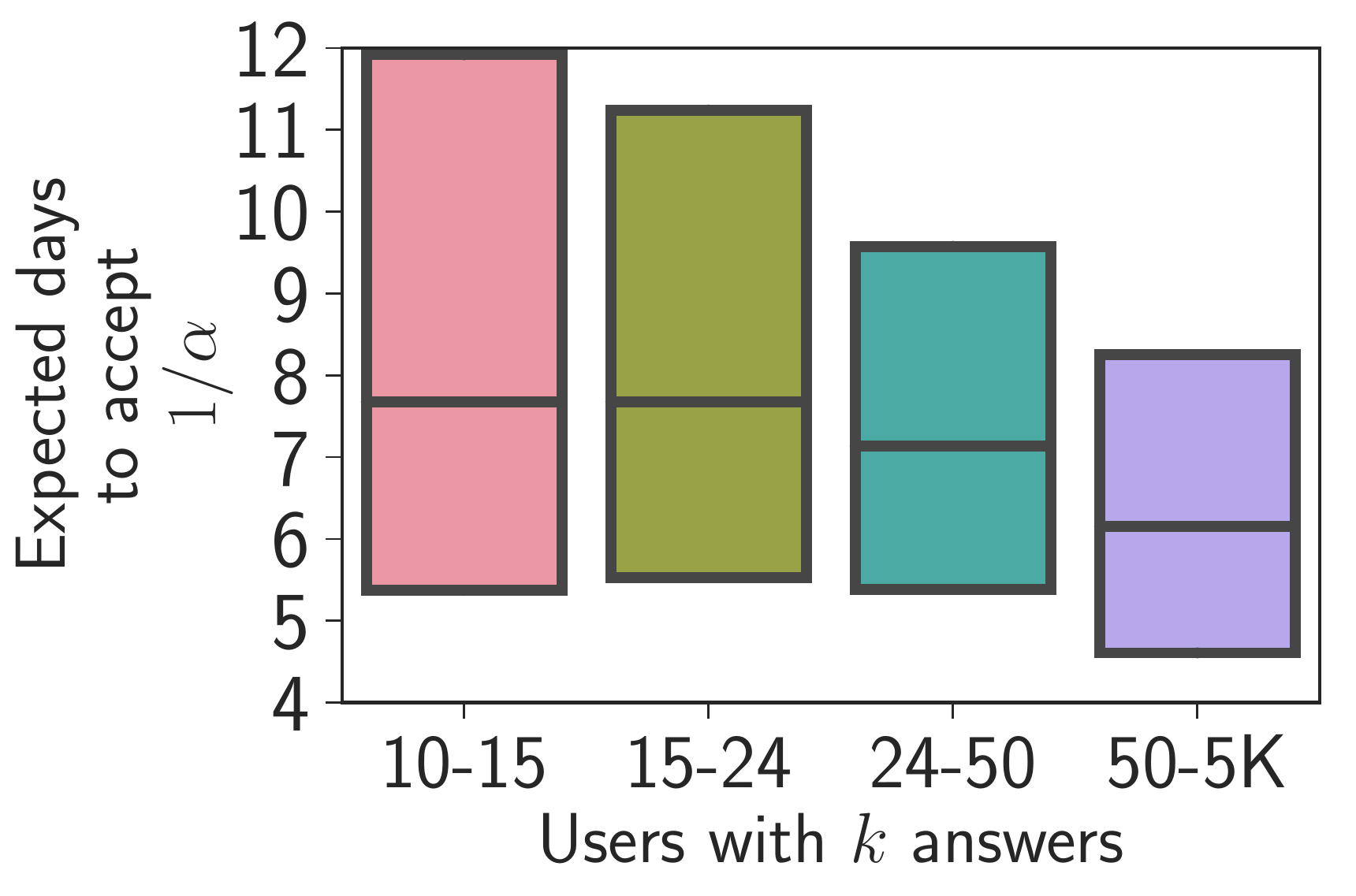} \\
  \small \hspace{6pt} Music & \small \hspace{6pt} Politics & \small \hspace{12pt} \textit{Stack Overflow} 
  \end{tabular}}
     \subfloat[Analysis of parameter $\gammab$]{\setlength\tabcolsep{3pt}
  \begin{tabular}{c c c }
  \includegraphics[width=0.15\textwidth]{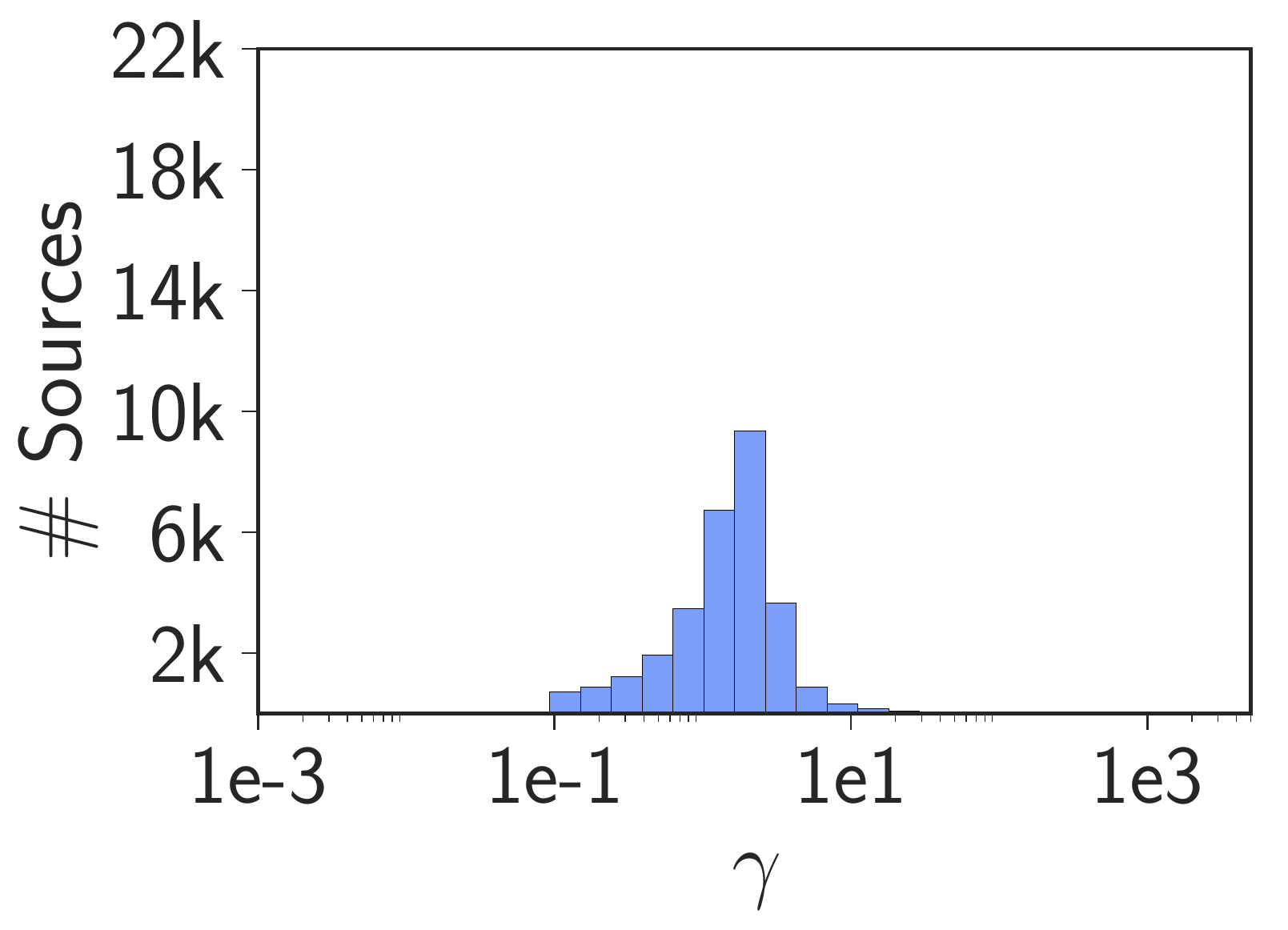} &
  \includegraphics[width=0.15\textwidth]{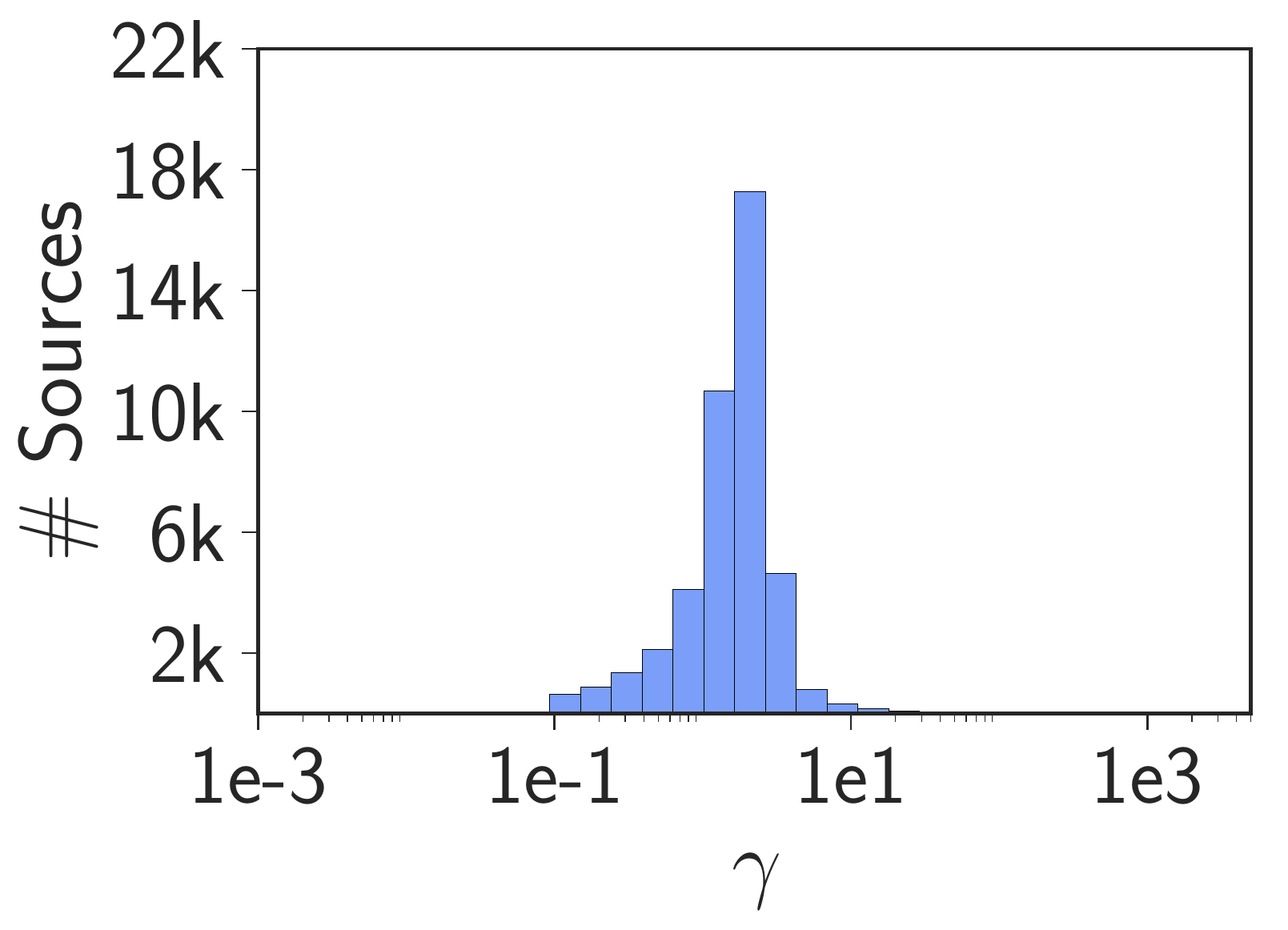} &
  \includegraphics[width=0.15\textwidth]{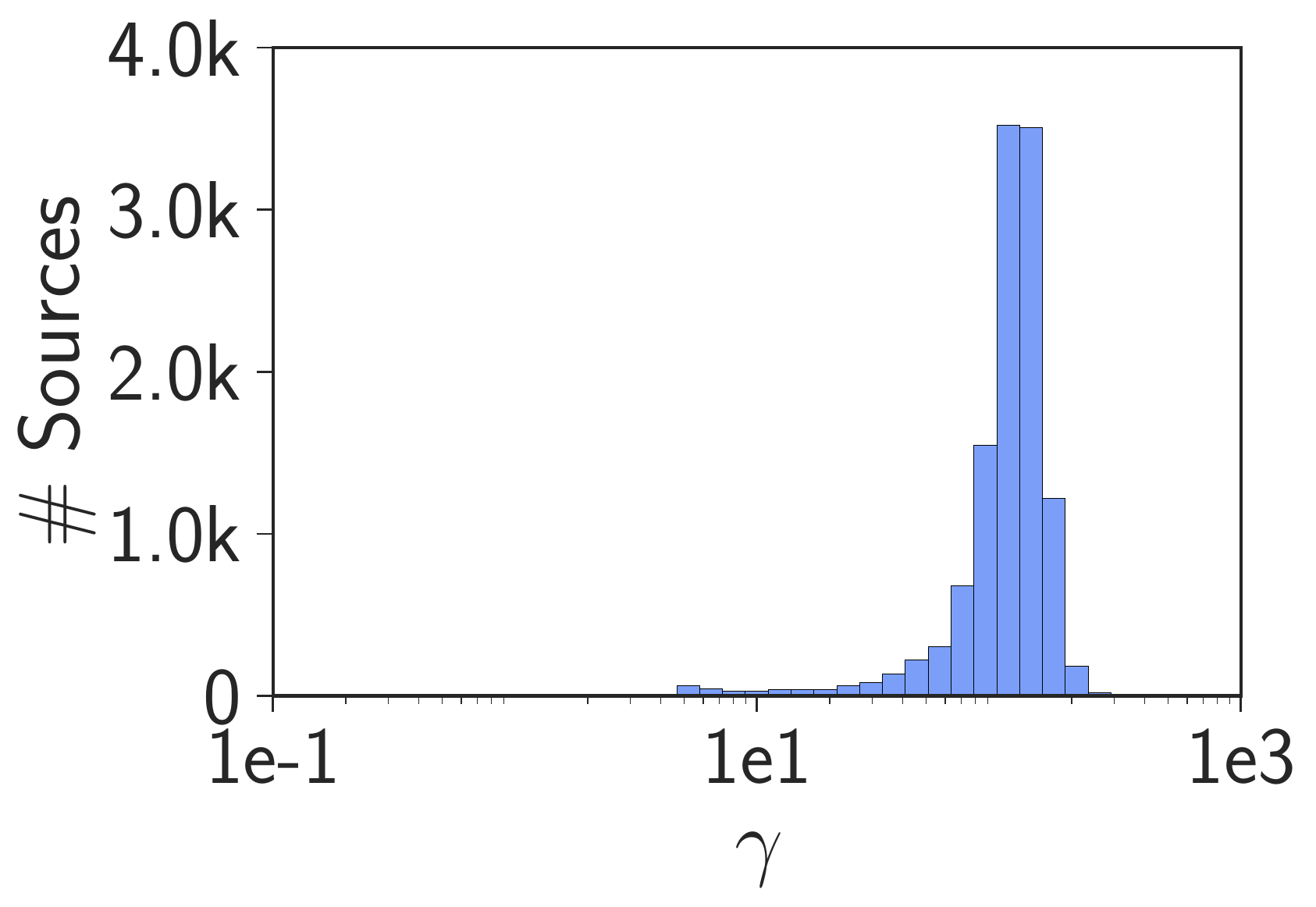} \\
  \includegraphics[width=0.15\textwidth]{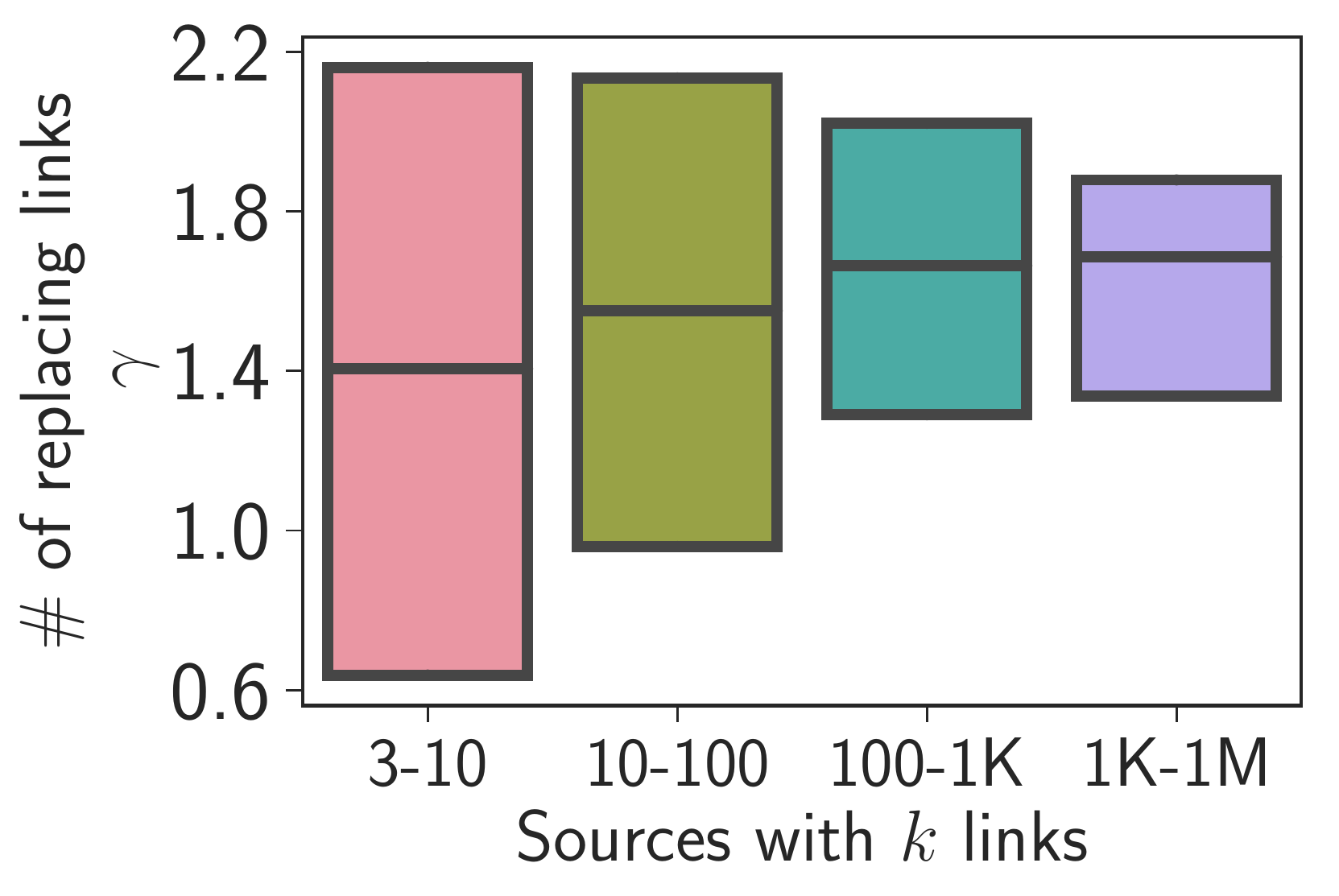} &
  \includegraphics[width=0.15\textwidth]{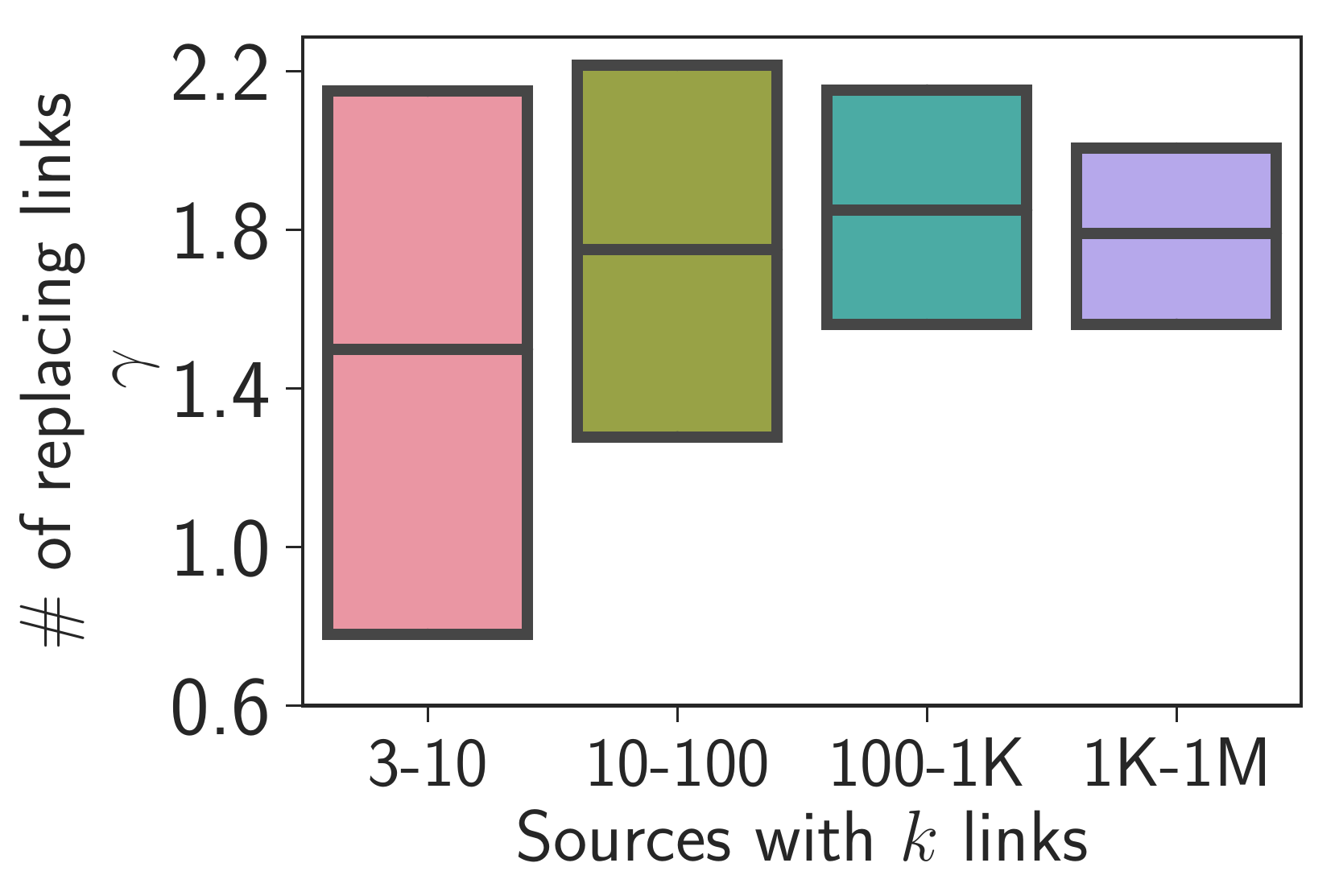} &
  \includegraphics[width=0.15\textwidth]{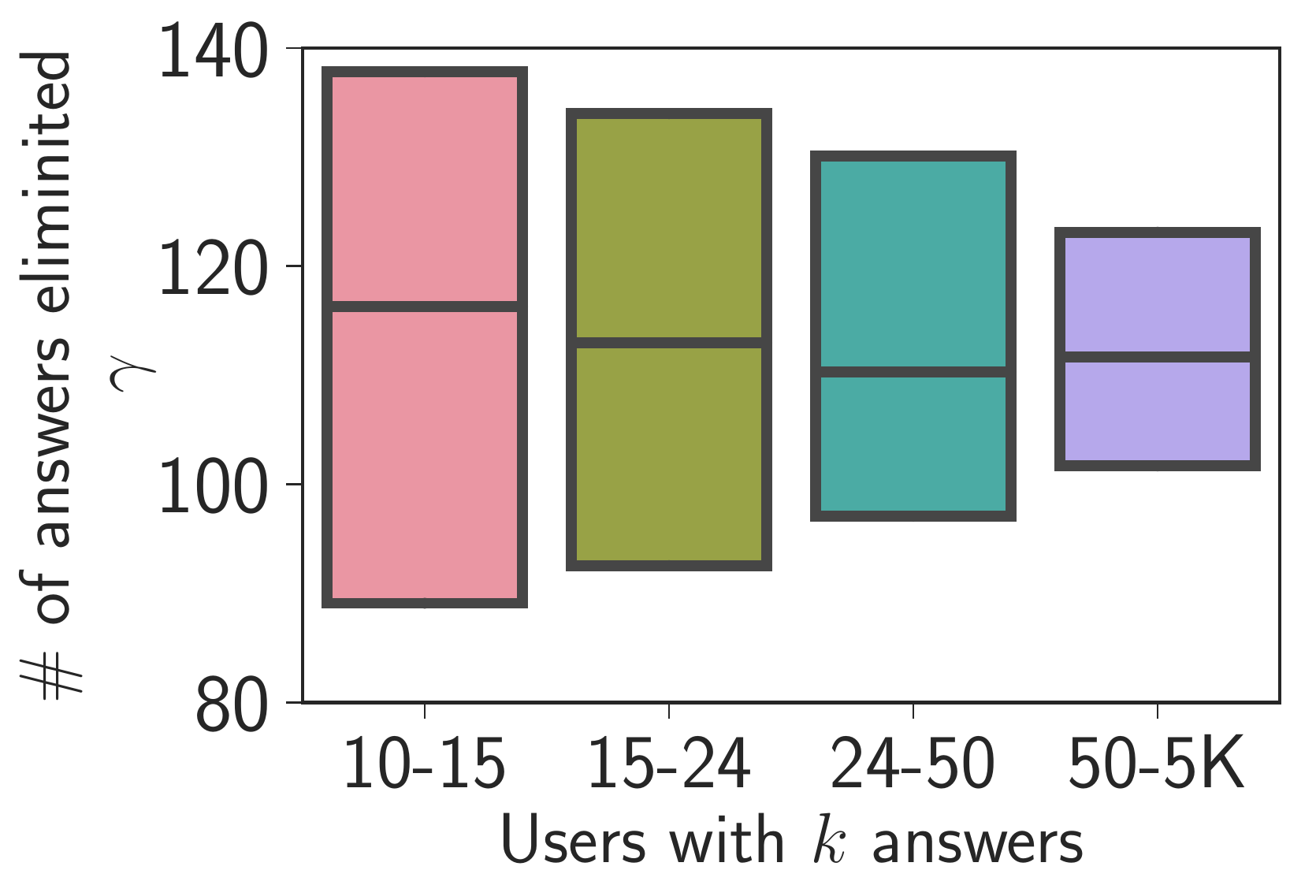} \\
  \small \hspace{6pt} Music & \small \hspace{6pt} Politics & \small \hspace{10pt} \textit{Stack Overflow}
  \end{tabular}}

  \caption{Source Trustworthiness. Panels (a) and (b) show the distributions of the parameters $\alphab$ and $\gammab$ for the Web sources in \textit{Wikipedia} for the topics
  ``music'' and ``politics''  and for the \textit{Stack Overflow} users, respectively.
In both panels, the top row shows the distributions across all sources, while the bottom row shows the distributions for four set of sources, grouped by their popularity in the case
of  \textit{Wikipedia} and by the number of answered questions in the case of \textit{Stack Overflow} users.
In \textit{Wikipedia}, the evaluation mechanism is refutation and thus larger values of $1/\alphab$ correspond to more trustworthy users whose contributed content is refuted more rarely.
In \textit{Stack Overflow}, the evaluation mechanism is verification and thus smaller values of $1/\alphab$ correspond to more trustworthy users whose contributed content is verified quicker.
In both cases, higher values of $\gammab$ imply a larger impact on the overall reliability of the knowledge item (\ie, article and question) after an evaluation.}
\label{fig:unrel-replace}
\end{figure*}

\xhdrq{Can we predict which of the answers to a question in Stack Overflow will be accepted?}
Unlike Wikipedia where each article receives multiple evaluations (\ie, deleted links), we have only one evaluation (\ie, accepted answer) for every question in Stack Overflow. This property prevents us from
estimating question difficulty in the test set and subsequently making predictions similar to that of \textit{Wikipedia}. However, we can estimate users' reliability from all the questions in the training set and predict which of several competing answers to a question will be most likely verified.

\emph{--- Experimental setup:}
We first split all questions (and corresponding answers) into a training set (90\% of the questions) and test set (the remaining 10\%) at
random, then fit the parameters of the evaluation process given by Eq.~\ref{eq:lambda_survival} using only the evaluation
times of the answers in the training set, and finally predict which answers will be accepted in the test set by computing
the expected verification time for all answers to a question using the fitted model and selecting the earliest estimated verification time.
We compare the performance of our model with two baselines: ``Source'' and ``Correct Answers". ``Source'' only accounts for the trustworthiness
of the sources (users) and ignores the intrinsic reliability (difficulty) of the questions. Thus, it computes the expected verification time of an answer
in the test set as the average verification time of all the answers provided by its associated source user in the training set.
Then, for each question in the test set, this baseline selects the answer with the lowest expected verification time.
``Correct Answers" ranks sources (users) according to the number of accepted answers posted by each user in the training set.
Then, for each question in the test set, it selects the answer with the highest ranked associated source.

\emph{--- Results:}
Figure~\ref{fig:rocauc}(b) summarizes the results by means of success rate
for different training set sizes.
Note that, unlike in
the \textit{Wikipedia} experiment, this prediction task does not correspond to a binary classification problem and therefore AUC is not a suitable metric in this case.
Our model always achieves a rate of success over $0.47$, consistently beats both baselines and, as expected, it becomes more accurate as we feed more events into the estimation procedure.
Note that, for most questions, there are more than two answers and the success rate of a random baseline is $0.41$.
The above results suggest that one needs to account for both the users'{} trustworthiness and the difficulty of the questions to be able to accurately predict which answer will be accepted,
in agreement with previous work~\cite{anderson16kdd}.

\xhdrq{Do our model parameters provide a meaningful and interpretable measure of source trustworthiness?}
We answer this question by analyzing the source parameters $\gammab_{s}$ and $\alphab_{s}$ estimated by our parameter estimation method, both in \textit{Wikipedia}
and \textit{Stack Overflow}.

First, we pay attention to the $20$ most used web sources in \textit{Wikipedia} for two topics, \ie, politics and music, and active users in
\textit{Stack Overflow} with over $100$ answers, and rank them in terms of source trustworthiness (\ie, in \textit{Wikipedia}, higher trustworthiness means lower $\alpha_s$, while in
\textit{Stack Overflow} higher trustworthiness means higher $\alpha_s$).
Then, we compute the probability that a statement supported by each source is refuted in less than $6$ months in \textit{Wikipedia} or ve\-ri\-fied in less than
$4$ days in \textit{Stack Overflow} due to only the source trustworthiness (\ie, setting  $\betab = 0$).
Table ~\ref{fig:unrel-wiki-table} and~\ref{fig:unrel-so-table} summarize the results, which reveal several interesting pa\-tterns.
For example, our model identifies social networking sites such as Twitter, which often accumulate questionable facts and opinionated information,
as untrustworthy sources for music in \textit{Wikipedia}. Similarly, for articles related to politics, some notable news agencies close to the left of the
political spectrum are considered to be more trustworthy, in agreement with previous studies on political bias in Wikipedia~\cite{greenstein2012wikipedia}.
Moreover, users with high reputation, as computed by \textit{Stack Overflow} itself, are indeed identified in our framework as trustworthy. However, the ranking
among these users in terms of reputation does not always match our measure of trustworthiness since it also takes into account other factors such as number of up-votes on questions and answers.

Next, we look at the source parameters at an aggregate level by means of their empirical distribution across users.
Fi\-gure~\ref{fig:unrel-replace} summarizes the results, which show that: (i) the distributions are remarkably alike across both topics in \textit{Wi\-ki\-pe\-dia} and
(ii) $\gamma$ values are distributed similarly both for \textit{Stack Overflow} and \textit{Wikipedia}, however, $\alpha$ values are distributed differently since
they capture a different mechanism, verification instead of refutation.
Finally, we group web sources in \textit{Wikipedia} by popularity and users of \textit{Stack Overflow} by number of contributed answers, and analyze the source
parameters. We summarize the results in Figure~\ref{fig:unrel-replace}, which show that:
(i) more popular web sources in \textit{Wikipedia} and more active users in \textit{Stack Overflow} tend to be more trustworthy, \ie, lower (higher) $\alpha$ in
\textit{Wikipedia} (\textit{Stack Overflow});
(ii)  popular sources in \textit{Wikipedia} have a larger impact on the reliability of the article, triggering a larger number of new statements additions (\ie, larger
values of $\gamma$) after a refutation;
and, (iii) there is ample variation across sources in terms of trustworthiness within all groups.

\begin{figure}[t]
    \centering
    \setlength\tabcolsep{1pt}
\begin{tabular}{c c}
   \includegraphics[width=0.40\textwidth]{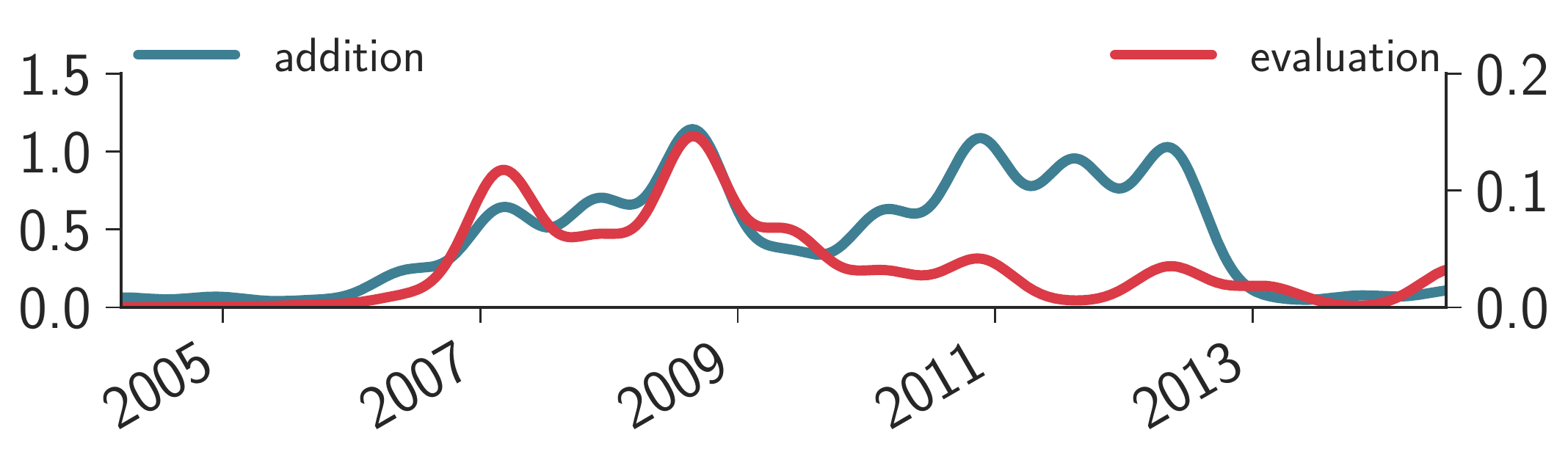}&      \hspace{10mm}
   \includegraphics[width=0.40\textwidth]{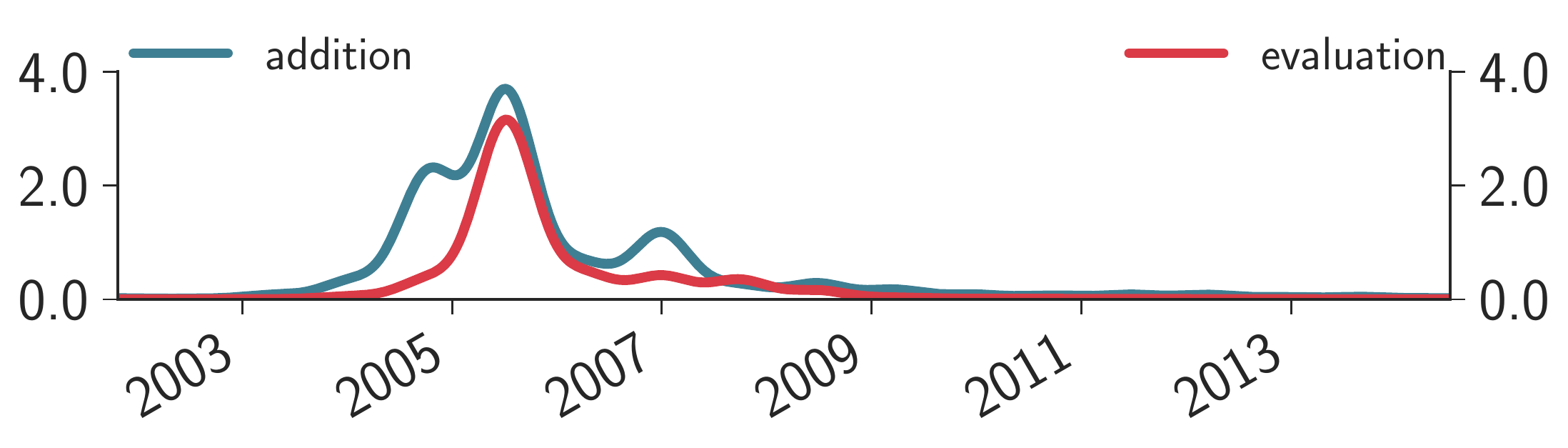} \\
   (a) Barack Obama'{}s biography & (b) George W. Bush'{}s biography \\
   \includegraphics[width=0.40\textwidth]{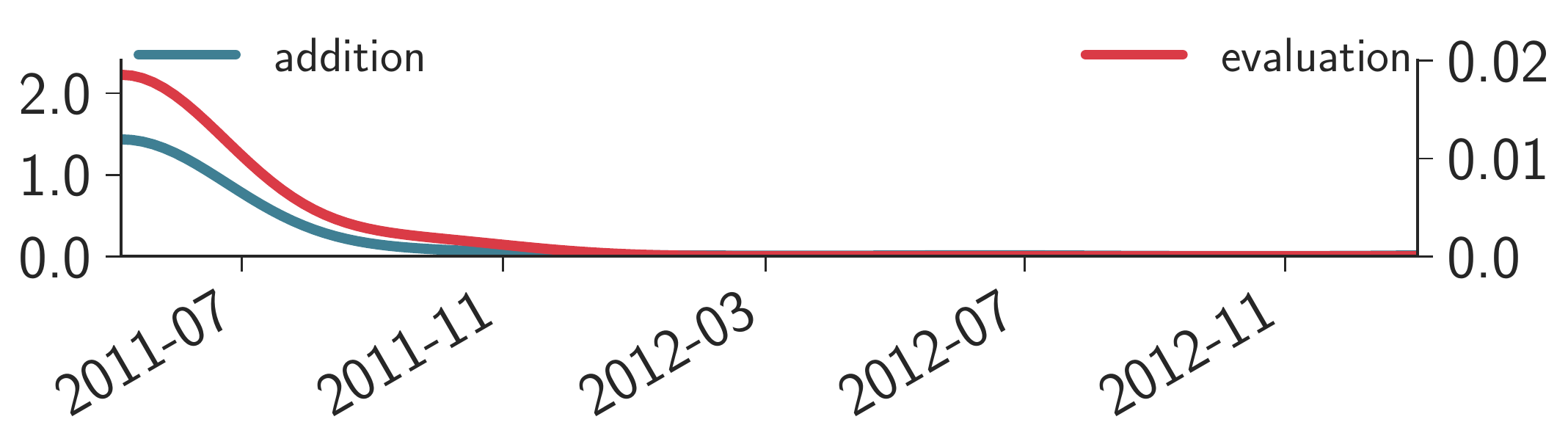} &      \hspace{10mm}
   \includegraphics[width=0.40\textwidth]{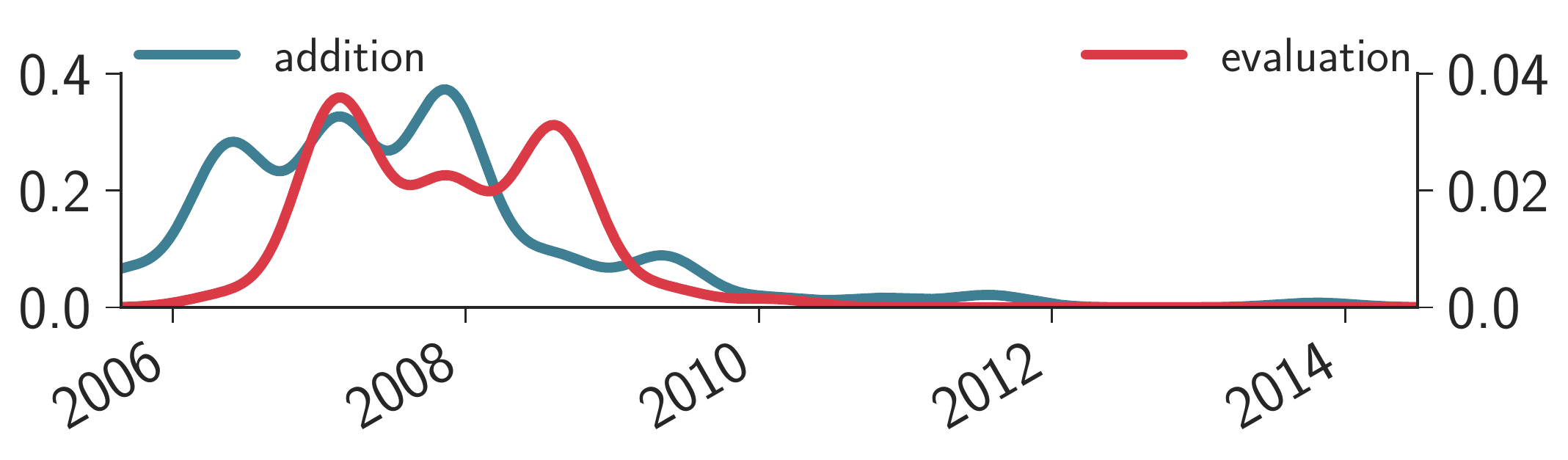}\\
   (c) ``2011 military intervention" in Libya & (d) TV show Prison Break  \\
\end{tabular}
\caption{Temporal evolution of the article intrinsic reliability for four \textit{Wikipedia} articles. The blue (red) line shows intensity of statement addition (evaluation) process.
Changes on the intrinsic reliability closely match external noteworthy events, often controversial, related to the corresponding article.}
\label{fig:docint-time}
\end{figure}

\xhdrq{What do the temporal evolution of the intrinsic reliability of Wikipedia articles tell us?}
In this section, we show that changes on the intrinsic reliability of a \textit{Wikipedia} article closely match external noteworthy events, often controversial,
related to the article.

Figure~\ref{fig:docint-time} shows the intrinsic reliability both in the statement addition process (first term in Eq.~\ref{eq:intensity-addition}), which captures the
arrival of new information, and the verification process (first term in Eq.~\ref{eq:lambda_survival}), which captures the controversy of the article, for four different articles -- Barack Obama'{}s biography,\footnote{\scriptsize \url{https://en.wikipedia.org/wiki/Barack$\_$Obama}} George W. Bush'{}s
biography,\footnote{\scriptsize \url{https://en.wikipedia.org/wiki/George$\_$W.$\_$Bush}} an article on 2011 military intervention
in Libya,\footnote{\scriptsize \url{https://en.wikipedia.org/wiki/2011$\_$military$\_$intervention$\_$in$\_$Libya}} and an article
on the TV show Prison Break.\footnote{\scriptsize \url{https://en.wikipedia.org/wiki/Prison$\_$Break}}
Each of the articles exhibits different characteristic temporal patterns.
In the two biographical articles and the article on the TV show, we find several peaks in the arrival of new information
and controversy over time, which typically match remarkable real-world events.
For example, in Barack Obama'{}s article, the peaks in early 2007 and mid-2008 coincide with the time in which he won the Democratic
nomination and the 2008 US election campaign; and, in the Prison Break'{}s article, the peaks coincide with the broadcasting of the four
seasons.
In contrast, in the article about 2011 military intervention in Libya, we only find one peak, localized at the beginning of the article life cycle,
which is followed by a steady decline in which the controversy lasts for a few months longer than the arrival of new information.
A comparison of the temporal patterns of new information arrivals and controversy within an article reveals a more subtle phenomenon:
while sometimes a peak in the arrival of new information also results in a peak of controversy, there are peaks in the arrival that do not
trigger controversy and vice-versa.

\xhdrq{What do the intrinsic reliability of Stack Overflow questions tell us?}
We answer this question by analyzing the parameters $\beta_d$ and $\phi_d$ estimated by our parameter estimation method for questions
in \textit{Stack Overflow}.
For each question, such parameters are unidimensional since, unlike \textit{Wikipedia}, the reliability of questions in \textit{Stack Overflow} does not typically
change over time.
Moreover, the parameters have natural interpretation: $\beta$ reflects the easiness of a question and $\phi$ reflects its popularity.

\begin{figure}
  \captionsetup[subfigure]{justification=centering}
  \centering
       \subfloat[][$P(\beta)$]{\includegraphics[width=0.30\textwidth]{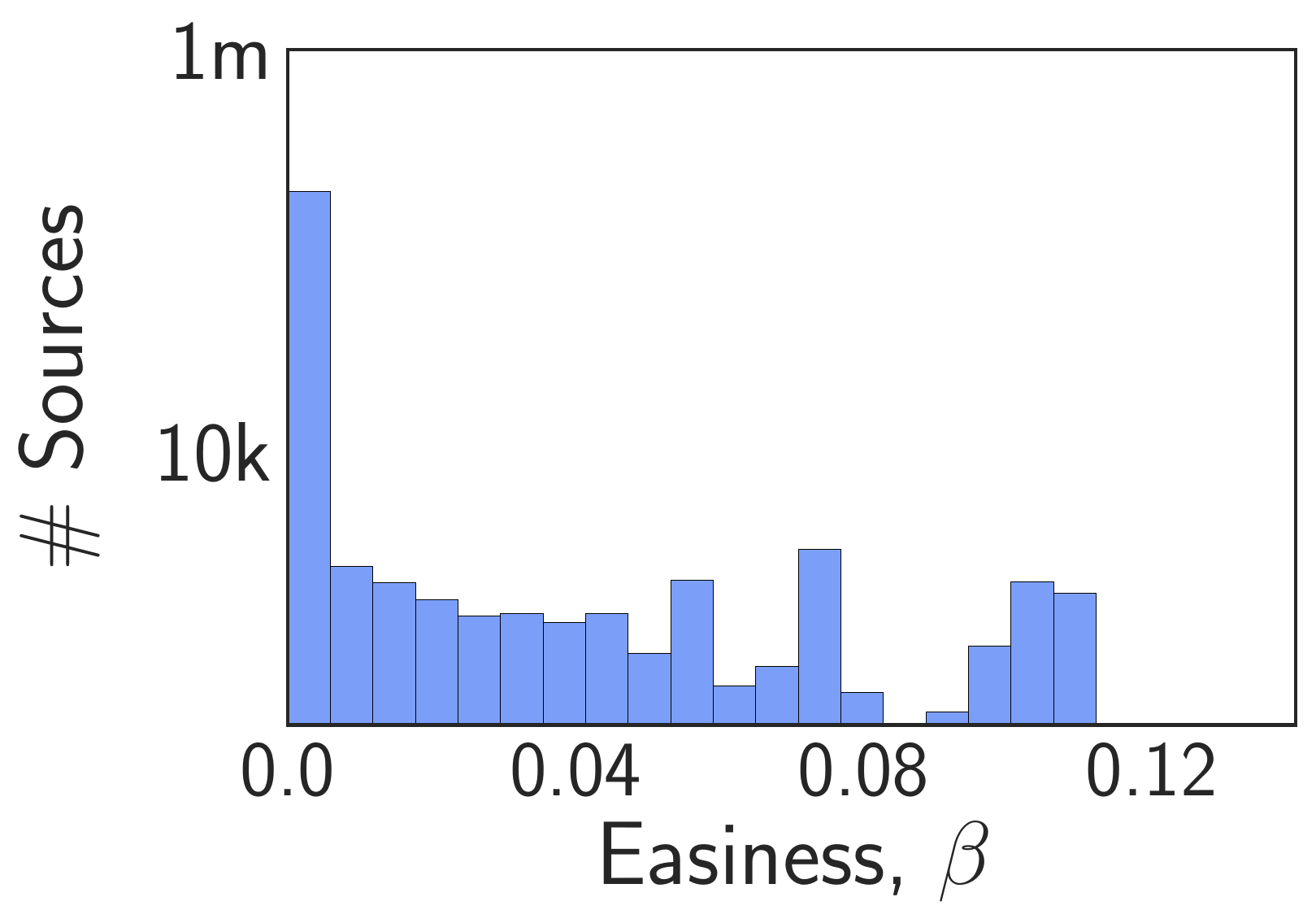}}\hspace{5mm}
       \subfloat[][$P(\phi)$]{\includegraphics[width=0.30\textwidth]{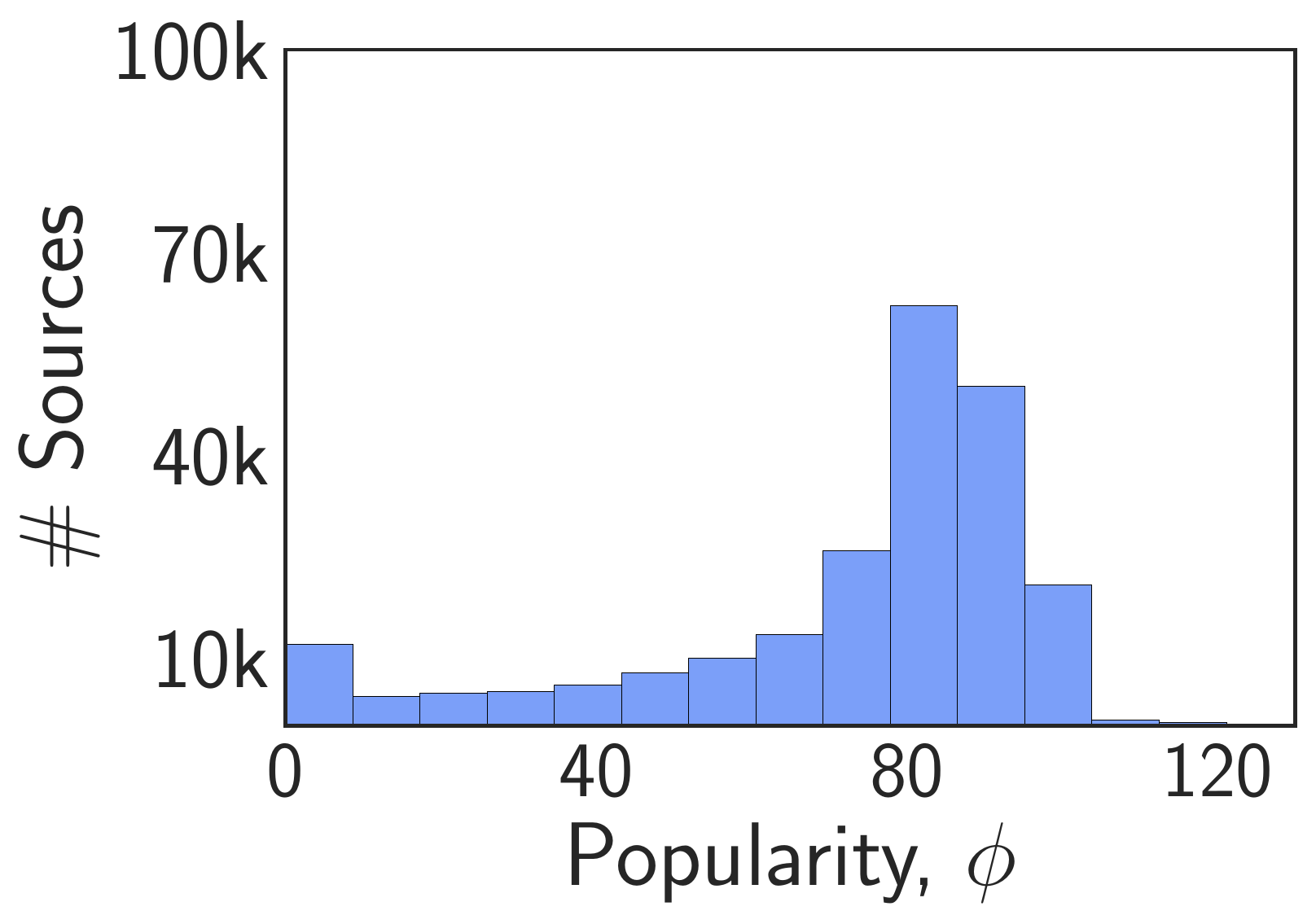}} \hspace{5mm}
       \subfloat[][$P(\phi, \beta)$]{\includegraphics[width=0.30\textwidth]{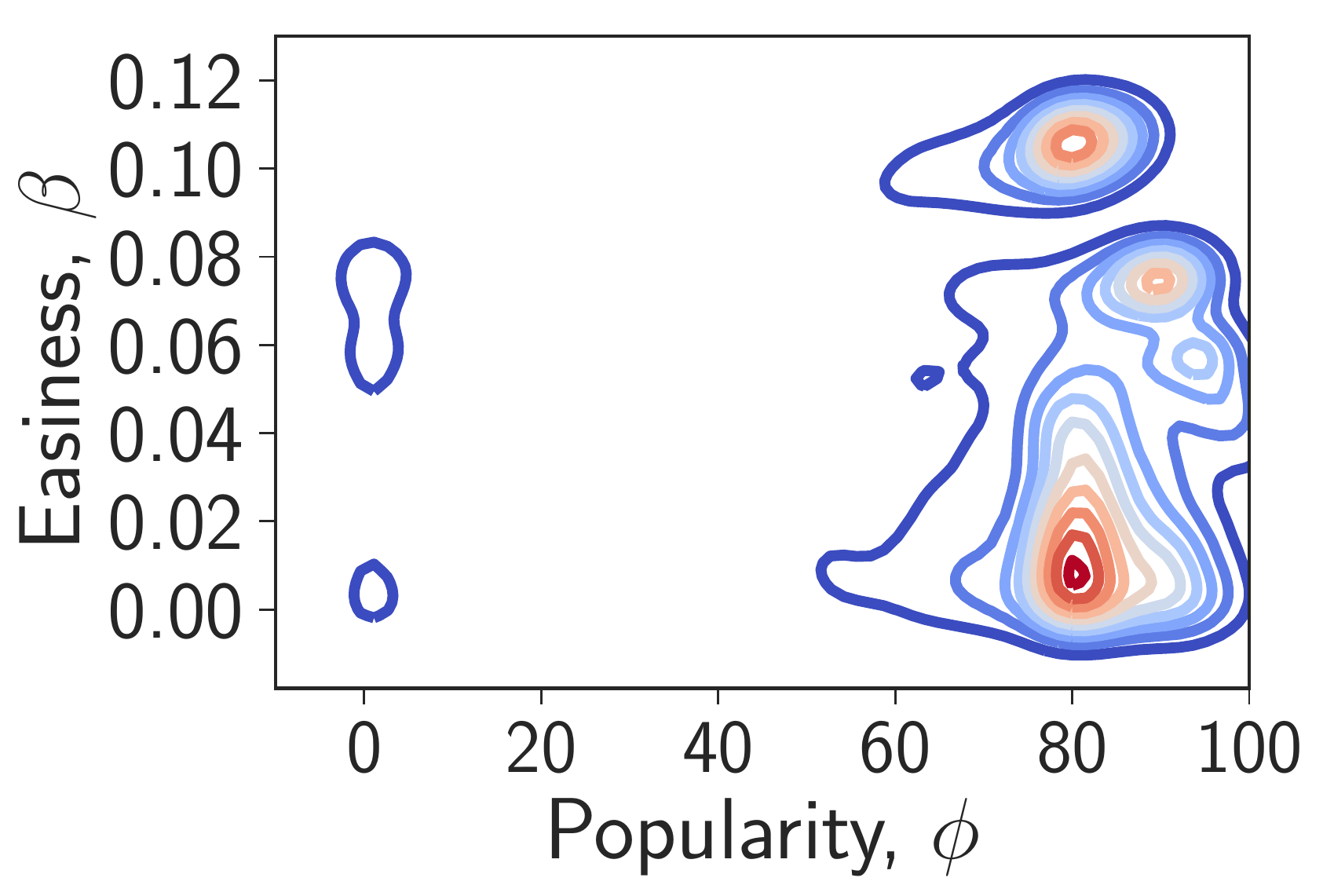}}
       \caption{Difficulty vs. popularity in \textit{Stack Overflow} questions. Panels (a) and (b) show the distribution of the parameters $\beta$ and $\phi$, which represent respectively the difficulty and the popularity of \textit{Stack Overflow} questions. Panel (c) shows the joint distribution of both parameters $\beta$ and $\phi$. Higher value of $\beta$ ($\phi$) implies easier (more popular) questions. } \label{fig:question_hist}
\end{figure}

Figures~\ref{fig:question_hist}(a-b) show the empirical marginal distribution of the parameters across questions and Figure~\ref{fig:question_hist}(c) shows the
joint distribution for questions with $\beta > 0$.
The results reveal four clusters: questions which are popular and easy, questions which are popular but difficult, questions that are not popular and difficult, and
questions that are not popular but easy.

\section{Conclusion}
\label{sec:conclusions}
In this paper, we proposed a temporal point process mo\-de\-ling framework of refutation and verification in online know\-ledge repositories and developed an efficient
convex optimization procedure to fit the parameters of our framework from historical traces of the refutations and verifications provided by the users of a knowledge
repository.
Then, we experimented with real-world data gathered from \textit{Wikipedia} and  \textit{Stack Overflow} and showed that our framework accurately predicts refutation
and verification events, provides an interpretable measure of information reliability and source trustworthiness, and yields interesting insights about real-world events.

Our work also opens many interesting directions for future work. For example, natural follow-ups to potentially improve the expressiveness of our modeling framework
include:
\begin{itemize}
\denselist
\item[1.] Consider sources can change their trustworthiness over time due to, \eg, increasing their expertise~\cite{learning17wsdm}.

\item[2.] Allow for non-binary refutation and verification events, \eg, partial refutations, ratings.

\item[3.] Augment our model to consider the trustworthiness of the user who refutes or verifies a statement.
\item[4.] Reduce number of parameters in the model by clustering sources and knowledge items.
\end{itemize}

Moreover, we experimented with data gathered from \textit{Wi\-ki\-pe\-dia} and \textit{Stack Overflow}, however, it would be interesting to apply our model (or augmented versions of our
model) to other knowledge repositories (\eg, \textit{Quora}), other types of online collaborative platforms (\eg, \textit{Github}), and the Web at large.
Finally, one can think of using our measure of trustworthiness, as inferred by our estimation method, to perform credit assignment in online collaborative platforms---in \textit{Wikipedia},
one could use our model to identify trustworthy users (or dedicated editors) who can potentially make an article more reliable and stable.

\section{Acknowledgments}
\label{sec:acknowledgments}
Authors would like to thank Martin Thoma for his comments in improving the
quality of paper.


\bibliographystyle{abbrv}
\bibliography{refs}

\end{document}